\documentclass[12pt,preprint]{emulateapj}

\usepackage{lscape}

\shorttitle{COSMOS AGN}
\shortauthors{Trump et~al.}

\begin{document}

\title{The COSMOS AGN Spectroscopic Survey I: XMM Counterparts\footnotemark[1]}

\author{
  Jonathan R. Trump\footnotemark[2], 
  Chris D. Impey\footnotemark[2], 
  Martin Elvis\footnotemark[3], 
  Patrick J. McCarthy\footnotemark[4], 
  John P. Huchra\footnotemark[3], 
  Marcella Brusa\footnotemark[5], 
  Mara Salvato\footnotemark[6], 
  Peter Capak\footnotemark[6], 
  Nico Cappelluti\footnotemark[5], 
  Francesca Civano\footnotemark[3], 
  Andrea Comastri\footnotemark[7], 
  Jared Gabor\footnotemark[2], 
  Heng Hao\footnotemark[3], 
  Gunther Hasinger\footnotemark[5], 
  Knud Jahnke\footnotemark[8], 
  Brandon C. Kelly\footnotemark[2], 
  Simon J. Lilly\footnotemark[9], 
  Eva Schinnerer\footnotemark[8], 
  Nick Z. Scoville\footnotemark[4], 
  and Vernesa Smol\v{c}i\'{c}\footnotemark[4]
}

\footnotetext[1]{ 
  Based on observations with the NASA/ESA \emph{Hubble Space
  Telescope}, obtained at the Space Telescope Science Institute, which
  is operated by AURA Inc, under NASA contract NAS 5-26555; the
  XMM-Newton, an ESA science mission with instruments and
  contributions directly funded by ESA Member States and NASA; the
  Magellan Telescope, which is operated by the Carnegie Observatories;
  and the MMT, operated by the MMT Observatory, a joint venture of the
  Smithsonian Institution and the University of Arizona.
\label{cosmos}}

\footnotetext[2]{
   Steward Observatory, University of Arizona, 933 North Cherry
   Avenue, Tucson, AZ 85721
\label{Arizona}}

\footnotetext[3]{
  Harvard-Smithsonian Center for Astrophysics, 60 Garden Street,
  Cambridge, MA 02138
\label{CfA}}

\footnotetext[4]{
  Observatories of the Carnegie Institute of Washington, Santa Barbara
  Street, Pasadena, CA 91101
\label{Carnegie}}

\footnotetext[5]{
  Max Planck-Institut f\"ur Extraterrestrische Physik,
  Giessenbachstrasse 1, D-85748 Garching, Germany
\label{Max Planck}}

\footnotetext[6]{
  California Institute of Technology, MC 105-24, 1200 East California
  Boulevard, Pasadena, CA 91125
\label{Caltech}}

\footnotetext[7]{
  INAF - Osservatorio Astronomico di Bologna, via Ranzani 1, 40127
  Bologna, Italy
\label{inaf}}

\footnotetext[8]{
  Max Planck Institut f\"ur Astronomie, K\"onigstuhl 17, D-69117
  Heidelberg, Germany
\label{Max Planck astro}}

\footnotetext[9]{
  Department of Physics, ETH Zurich, CH-8093 Zurich, Switzerland
\label{Zurich}}

\def\etal{et al.}

\newcommand{\FeII}{\hbox{{\rm Fe}\kern 0.1em{\sc ii}}}
\newcommand{\FeIII}{\hbox{{\rm Fe}\kern 0.1em{\sc iii}}}
\newcommand{\Ha}{\hbox{{\rm H}\kern 0.1em$\alpha$}}
\newcommand{\Hb}{\hbox{{\rm H}\kern 0.1em$\beta$}}
\newcommand{\MgII}{\hbox{{\rm Mg}\kern 0.1em{\sc ii}}}
\newcommand{\CIV}{\hbox{{\rm C}\kern 0.1em{\sc iv}}}
\newcommand{\CIII}{\hbox{{\rm C}\kern 0.1em{\sc iii}]}}
\newcommand{\OII}{\hbox{{\rm [O}\kern 0.1em{\sc ii}]}}
\newcommand{\OIII}{\hbox{{\rm [O}\kern 0.1em{\sc iii}]}}
\newcommand{\NII}{\hbox{{\rm [N}\kern 0.1em{\sc ii}]}}
\newcommand{\NeIV}{\hbox{{\rm [N}\kern 0.1em{\sc iv}]}}

\begin{abstract}

We present optical spectroscopy for an X-ray and optical flux-limited
sample of 677 XMM-Newton selected targets covering the 2 deg$^2$
COSMOS field, with a yield of 485 high-confidence redshifts.  The
majority of the spectra were obtained over three seasons (2005-2007)
with the IMACS instrument on the Magellan (Baade) telescope.  We also
include in the sample previously published Sloan Digital Sky Survey
spectra and supplemental observations with MMT/Hectospec.  We detail
the observations and classification analyses.  The survey is 90\%
complete to flux limits of $f_{\rm 0.5-10 keV}>8 \times 10^{-16}~{\rm
erg~cm^{-2}~s^{-1}}$ and $i_{\rm AB}^+<22$, where over 90\% of targets
have high-confidence redshifts.  Making simple corrections for
incompleteness due to redshift and spectral type allows for a
description of the complete population to $i_{\rm AB}^+<23$.  The
corrected sample includes 57\% broad emission line (Type 1,
unobscured) AGN at $0.13<z<4.26$, 25\% narrow emission line (Type 2,
obscured) AGN at $0.07<z<1.29$, and 18\% absorption line
(host-dominated, obscured) AGN at $0<z<1.22$ (excluding the stars that
made up 4\% of the X-ray targets).  We show that the survey's limits
in X-ray and optical flux include nearly all X-ray AGN (defined by
$L_{\rm 0.5-10 keV}>3 \times 10^{42}$ erg~s$^{-1}$) to $z<1$, of both
optically obscured and unobscured types.  We find statistically
significant evidence that the obscured to unobscured AGN ratio at
$z<1$ increases with redshift and decreases with luminosity.


\end{abstract}


\section{Introduction}

Active Galactic Nuclei (AGN) are the brightest persistent
extragalactic sources in the sky across nearly all of the
electromagnetic spectrum.  Only in the relatively narrow range of
infrared (IR) through ultraviolet (UV) wavelengths are AGN often
outshone by stellar emission.  Here the central engines can be dimmed
by obscuring dust and gas while starlight, either direct or absorbed
and re-emitted by dust, peaks.  Historically, the largest AGN surveys
have been based on optical selection \citep[e.g.][]{sch83, hew95, hes,
cro01, schn07}.  Yet in both the local and distant universe, obscured
AGN are generally thought to outnumber their unobscured counterparts
\citep[e.g. ][]{mai95, gil01, ste04, bar05, mar05, dad07, trei08},
indicating that optical surveys probably miss the majority of AGN.  A
more complete census of AGN must use their X-ray, mid-infrared, and
radio emission, where obscuration and host contamination are
minimized.  X-ray and mid-IR selected surveys do in fact reveal a far
greater space density of AGN than optical selection: for example, the
{\it Chandra} deep fields reveal AGN sky densities 10-20 times higher
than those of optically selected surveys to the same limiting optical
magnitudes \citep{bau04, ris04, bra05}.  However, most X-ray and
mid-IR surveys either have significantly smaller areas and numbers of
AGN or are wide-area but substantially shallower than optical surveys
\citep[e.g.][]{sch00, lon03}.  Here we present a deep spectroscopic
survey of AGN both without the biases of optical selection and over a
relatively large field.


The Cosmic Evolution Survey \citep[COSMOS,][]{sco07}\footnote{The
COSMOS website is http://cosmos.astro.caltech.edu/.} is built upon an
HST Treasury project to fully image a 2 deg$^2$ equatorial field.  The
590 orbits of HST ACS $i$-band observations have been supplemented by
observations at wavelengths from radio to X-ray, including deep VLA,
Spitzer, CFHT, Subaru (6 broad bands and 14 narrow bands), GALEX,
XMM-Newton, and Chandra data.  Here we present a complete
spectroscopic survey of XMM-selected AGN in the COSMOS field.  Most
(601) targets have spectra taken with the IMACS spectrograph
\citep{big98} on the Magellan telescope, including 282 spectra
previously published by \citep{tru07}.  An additional 76 X-ray targets
were excluded from IMACS observations because they already had SDSS
spectra.  For 134 of the targets with IMACS coverage, we additionally
acquired spectra with the Hectospec spectrograph \citep{fab05} on the
MMT telescope as ancillary data with extended blue coverage.  In
total, we were able to target 52\% (677/1310) of the available $i_{\rm
AB}^+<23.5$ X-ray targets, resulting in 485 high-confidence redshifts.
The relevant observing strategies and configurations are described in
detail in \S 2.  We were 90\% in assigning high-confidence redshifts
to all spectral types at $i_{\rm AB}^+<22$, with decreasing
confidence, dependent on both redshift and spectral type, at fainter
magnitudes.  The IMACS spectroscopy campaign additionally targeted AGN
candidates selected by their radio (VLA, 605 targets) and IR
(Spitzer/IRAC, 236 targets) emission, but these objects are not
included in this study and will be presented in future work.

We place this work in the context of other large X-ray AGN surveys in
Figure \ref{fig:surveys}, where the left panel compares the X-ray
depth, areal coverage, and number of sources for various X-ray AGN
surveys.  The right panel of Figure \ref{fig:surveys} shows our flux
limits with the customary ``AGN locus'' \citep{mac88}.  The depth of
XMM-Newton in COSMOS most closely resembles the AEGIS \citep{dav07}
survey, with roughly the same number of X-ray targets in both despite
their slight differences in area and X-ray depth.  There exists no
purely optical survey to the depth of our spectroscopy ($i_{\rm
AB}^+<23.5$) with this number of spectroscopic redshifts.  The AGN
spectroscopic campaign presented here is significantly deeper than
large optical surveys like the 2dF Quasar Redshift Survey
\citep{cro01} and the Sloan Digital Sky Survey \citep{schn07}.  In
particular, we present targets $\sim$60 times fainter than the main
SDSS spectroscopy ($g<19.1$), and $\sim$20 times fainter than the
deepest SDSS spectroscopy ($g<20.2$) for quasars, and our spectroscopy
reaches a (arbitrary) quasar/Seyfert boundary of $M_i=-23$ at $z \sim
3$.  Surveys like the VIMOS Very Deep Survey \citep[VVDS,][]{gav06}
may reach similarly faint magnitudes ($i \lesssim 24$ in VVDS) but
have far fewer AGN (130 in VVDS).  We additionally note that the
Magellan AGN sample will eventually be augmented by $\sim$300 X-ray
AGN from the faint zCOSMOS survey of galaxy redshifts with VLT/VIMOS
\citep{lil08}.

We discuss the analysis of the spectra in \S 3, including the methods
for classifying the AGN and determining redshifts.  In \S 4 we
characterize the completeness of the survey and discuss the
populations of different AGN types.  We use the sample to understand
the X-ray AGN population in \S 5, and we discuss future projects using
this dataset in \S 6.  We adopt a cosmology consistent with WMAP
results \citep{spe03} of $h=0.70$, $\Omega_M=0.3$,
$\Omega_{\Lambda}=0.7$.

Throughout the paper we use ``unobscured'' to describe Type 1 AGN with
broad emission lines and ``obscured'' to describe X-ray AGN where the
host galaxy light dominates the optical continuum.  Thus we use
``obscured AGN'' to describe both spectroscopically-defined Type 2 AGN
(with narrow emission lines, classified as ``nl'' or ``nla'' in the
catalog) and XBONGs \citep[X-ray bright, optically normal galaxies,
classified as ``a'' in the catalog, see also][]{com02,rig06,civ07}.
It is important to note that our designation as ``obscured'' does not
necessarily describe the physical reason for the faint optical nuclear
emission: the AGN might simply be under-luminous in the optical
instead of being hidden by obscuring material.  Indeed, many Type 2
AGN appear to be unobscured in the X-rays \citep{szo04}, while broad
absorption line (BAL) Type 1 AGN are typically X-ray obscured
\citep{bra00, gal06}.  We also note that even our ``obscured'' AGN
types have moderate X-ray luminosity and we are not sensitive to
heavily X-ray obscured (e.g., Compton-thick, $N_H \gtrsim 1 \times
10^{24}$ cm$^-2$) AGN which are too faint for our XMM-Newton
observations.

\section{Observations}

\subsection{XMM}

The COSMOS field has been observed with XMM-{\it Newton} for a total
of $\sim 1.55$ Ms at the homogeneous vignetting-corrected depth of
$\sim 50$ ks \citep{has07, capp07, capp08}.  The final catalog
includes 1887 point-like sources detected in at least one of the soft
(0.5-2 keV), hard (2-10 keV) or ultra-hard (5-10 keV) bands down to
limiting fluxes of $5 \times 10^{-16}$, $3.3 \times 10^{-15}$, and $5
\times 10^{-15}$ erg cm$^{-2}$ s$^{-1}$, respectively \citep[see][for
more details]{capp07, capp08}.  The detection threshold corresponds to
a probability $<4.5 \times 10^{-5}$ that a source is instead a
background fluctuation.  The XMM fluxes have been computed converting
the count-rate into flux assuming a spectral index $\Gamma=2.0$ and
Galactic column density $N_H=2.5 \times 10^{20}$ cm$^2$ for 0.5-2 keV
and $\Gamma=1.7$ and Galactic column density $N_H=2.5 \times 10^{20}$
cm$^2$ for 2-10 keV.  Following \citet{bru08}, we exclude 24 sources
which are a blend of two {\it Chandra} sources and 26 faint XMM
sources coincident with diffuse emission \citep{fin08}.  We impose a
brighter flux limit than the full catalog because the XMM-{\it Newton}
observations were not complete until the 3rd season (2007) of
spectroscopic observing.  Figure \ref{fig:xmmsens} shows the X-ray
sensitivity for each of the three seasons of IMACS, revealing that the
first two seasons (2005-2006) suffer from slightly shallower X-ray
catalogs.  The sample we use is limited to flux limits of the 50\% XMM
coverage area, which has only 186 few sources than from the limits of
the entire XMM coverage.  The sample includes 1651 X-ray sources
detected at fluxes larger than 1$\times 10^{-15}$ cgs, 6$\times
10^{-15}$ cgs, 1$\times 10^{-14}$ cgs, in the 0.5-2 keV, 2-10 keV or
5-10 keV bands, respectively, as presented by \citet{bru08}.

\citet{bru08} associated the X-ray point sources with optical
counterparts using the likelihood ratio technique to match to the
optical, near-infrared (K-band) and mid-infrared (IRAC) photometric
catalogs \citep{cap07}.  The images for the XMM-COSMOS subsample
additionally covered by {\it Chandra} observations were matched to the
Chandra/ACIS images by visual inspection \citep{elv08, puc08, civ08}.
We use the COSMOS {\it Chandra} observations for reliability checks
only, since it covers only the central 0.8 deg$^2$ and is still
undergoing basic analyses.

Of the 1651 sources in the XMM-COSMOS catalog described above, 1465
sources have an unique/secure optical counterpart from the
multiwavelength analysis with a probability of misidentification of
$<1\%$.  For an additional 175 sources, there is a second optical
source with a comparable probability to be the correct counterpart.
Because the alternate counterpart shows comparable optical to IR
properties \citep[and comparable photometric redshifts,][]{sal08} to
the primary counterpart, the primary counterpart can be considered
statistically representative of the true counterpart for these 175
X-ray sources, and we include the primary counterparts in the target
sample.  Eleven sources (outside the Chandra area) remain unidentified
because they had no optical or infrared counterparts (i.e., their
optical/infrared counterparts were fainter than our photometry).  We
designated the 1310 optical counterparts with $i_{\rm AB}^+ \leq 23.5$
(from the CFHT) as the X-ray selected targets for the spectroscopic
survey.

\subsection{Magellan/IMACS}

The bulk of the spectroscopic data comes from observations with the
Inamori Magellan Areal Camera and Spectrograph \citep[IMACS,][]{big98}
on the 6.5 m Magellan/Baade telescope.  The IMACS field of view is
$22{\arcmin}30{\arcsec} \times 21{\arcmin}10{\arcsec}$ (with only 10\%
vignetting at the extreme chip edge), requiring 16 tiled pointings to
fully observe the entire 2 deg$^2$ COSMOS field as shown in Figure
\ref{fig:field}.  We observed these 16 pointings over the course of 26
nights (18 clear) through three years, as detailed in Table
\ref{tbl:imacsnights}.  The total exposure time for each pointing is
4-6 hours (shown in Table \ref{tbl:imacsnights} and Figure
\ref{fig:field}).  Henceforth we refer to each pointing by its number
in Table \ref{tbl:imacsnights} and Figure \ref{fig:field}.  We were
able to simultaneously observe 200-400 spectra per mask: generally
$\sim$40 of these were the X-ray targets described here (shown in the
last column of Table \ref{tbl:imacsnights}), and the additional slits
were ancillary targets to be described in future work.  We were
generally able to target $\sim$50\% of the available $i_{\rm AB}^+
\leq 23.5$ X-ray targets in each tiled IMACS field, or 601/1310 X-ray
targets over the 2 deg$^2$.

All IMACS spectra were obtained over the wavelength range of 5600-9200
\AA, with the Moon below the horizon and a mean airmass of 1.3.  We
used the 200 l/mm grism in the first year and a 150 l/mm grism
designed and constructed for COSMOS in the second and third years.
The lower-resolution 150 l/mm grism had a resolution element of 10\AA.
Since all observed broad line AGN had line widths $>1500$ km~s$^{-1}$
and all observed narrow line AGN had line widths $<1000$ km~s$^{-1}$,
the resolution of the grism was sufficient to distinguish broad and
narrow line AGN.  The gain in S/N from 200 l/mm to 150 l/mm was only
marginal, but the 150 l/mm grism allowed for a maximum of 400 slits
per mask, $\sim$35\% more than the maximum 300 slits per mask for the
200 l/mm grism.  The slits were $11\arcsec \times 1 \arcsec$ ($55
\times 5$ pixels), though only $5{\farcs}4 \times 1\arcsec$ of the
slit was cut, so that an extra adjacent $5{\farcs}6$ was reserved as
an ``uncut region'' to accommodate ``nod-and-shuffle'' observing (see
below).  We attempted to observe each mask for 5 or more hours, which
achieves high completeness of AGN redshifts at $i_{\rm AB}^+ \simeq
23$, although as Figure \ref{fig:field} shows this was not always
achieved.  We estimate the impact of the nonuniform spectroscopic
depth on the sample's completeness in \S 4.1.

We observed using the ``nod-and-shuffle'' technique, which allowed for
sky subtraction and fringe removal in the red up to an order of
magnitude more precisely than conventional methods.  The general
principles of nod-and-shuffle are described by \citet{gla01}, and our
approach is detailed in Appendix 1 of \citet{abr04}.  Briefly, we
began observing with the target objects offset from the vertical
center of the cut region, 1/3 of the way from the bottom to the top
(that is, $1{\farcs}8$ from the bottom slit edge, and $3{\farcs}6$
from the top edge of the cut region and the cut/uncut boundary).
After 60 seconds we closed the shutter, nodded the telescope by
$1{\farcs}8$ (9 pixels) along the slit, and shuffled the charge to the
reserved uncut region.  The object was then observed for 60 seconds in
the new position, 2/3 of the way from the bottom to the top of the cut
region ($3{\farcs}6$ from the bottom and $1{\farcs}8$ from the top).
We then closed the shutter, nodded back to the original position, and
shuffled the charge back onto the cut region on the mask.  This cycle
was repeated (typically 15-20 times) with the net result that the sky
and object had been observed for equal amounts of time on identical
pixels on the CCD.  Nod-and-shuffle worked well while the seeing was
$\lesssim 1\arcsec$, which was true for all observations.

To extract and sky-subtract individual 2D linear IMACS spectra, we
used the publicly available Carnegie Observatories System for
MultiObject Spectroscopy (with coincidentally the acronym ``COSMOS,''
written by A. Oemler, K. Clardy, D. Kelson, and G. Walth and publicly
available at http://www.ociw.edu/Code/cosmos).  We combined the two
nod positions in the nod-and-shuffle data, then co-added the
individual 2D exposures of each pointing while rejecting cosmic rays
as 4.5$\sigma$ outliers from the mean of the individual exposures.
Wavelength calibration was performed using an He/Ne/Ar arc lamp
exposure in each slit.  The 2D spectra were extracted to 1D
flux-calibrated spectra using our own IDL software, adapted from the
{\tt ispec2d} package \citep{ispec}.  While flux calibration used only
a single standard star at the center of the IMACS detector, we
estimate by eye that vignetting has $<10\%$ effect on the spectral
shape or throughput across the field, in agreement with the
predictions of the IMACS manual.

IMACS spectra can be contaminated or compromised in several ways,
including 0th and 2nd order lines from other spectra, bad pixels and
columns, chip gaps, poorly machined slits, and cosmic rays missed
during co-adding.  To eliminate these artifacts, we generated bad
pixel masks for all 1D spectra by visual inspection of the calibrated
1D and 2D data.  The nod-and-shuffle 2D data were especially useful
for artifact rejection: any feature appearing in only one of the two
nod positions is clearly an artifact.  Pixels designated as bad in the
mask were ignored in all subsequent analyses.

We show 10 examples of IMACS spectra in Figures \ref{fig:spec1} and
\ref{fig:spec2}.  These spectra are representative of the targets in
the survey.  Each of these spectra are smoothed by the 5-pixel
resolution element.  We discuss each object below, with the spectral
classification, confidences, and redshift algorithms detailed in \S 3.
Briefly, $z_{\rm conf}=3,4$ refer to high confidence and $z_{\rm
conf}=1,2$ are lower confidence guesses (but see also \S 3.1 for the
subtleties in confidence assignment).  All spectra are publicly
available on the COSMOS IRSA server
(http://irsa.ipac.caltech.edu/data/COSMOS/).

\begin{enumerate}
    \item COSMOS J095909.53+021916.5, $i_{\rm AB}^+=20.05$, $z=0.38$,
      $z_{\rm conf}=4$: This is a low redshift Type 1 Seyfert.  The
      emission lines are bright and easily identified.

    \item COSMOS J095752.17+015120.1, $i_{\rm AB}^+=21.00$, $z=4.17$,
      $z_{\rm conf}=4$: This is a high redshift Type 1 quasar.
      Ly$\alpha$ is especially prominent along other broad emission
      features, and so this redshift is very reliable.

    \item COSMOS J095836.69+022049.0, $i_{\rm AB}^+=23.04$, $z=1.19$,
      $z_{\rm conf}=4$: We classify this target as a hybrid ``bnl''
      object with both broad and narrow emission lines.  The narrow
      \OII~ line is evident above the noise and strong broad \MgII is
      also present.

    \item COSMOS J095756.77+024840.9, $i_{\rm AB}^+=19.60$, $z=1.61$,
      $z_{\rm conf}=3$: In this spectrum, a broad emission line is cut
      off by a detector chip gap.  Identifying the broad feature as
      MgII and the minor narrow emission line at $\sim$6375\AA~ as
      \NeIV~yields a good redshift, but we assign only $z_{\rm
      conf}=3$ because of the uncertainty from the chip gap position.

    \item COSMOS J100113.83+014000.9, $i_{\rm AB}^+=20.49$, $z=1.56$,
      $z_{\rm conf}=2$: The blue end of this spectrum lies on a chip
      gap, and much of the red end is corrupted by second order
      features from another bright spectrum on the mask.  Only one
      broad emission line is present, and so while the target is
      clearly a Type 1 AGN, the line could be either CIII] or MgII.
      The redshift solution is degenerate and we assign only $z_{\rm
      conf}=2$.

    \item COSMOS J095821.38+013322.8, $i_{\rm AB}^+=19.16$, $z=0.44$,
      $z_{\rm conf}=4$: This spectrum contains several bright emission
      lines, and is clearly identified as a ``nl'' class object.  This
      object has $L_{0.5-10~{\rm keV}} < 3 \times 10^{42}$ and $-2 \le
      \log{f_X/f_O} \le -1$, and it is probably a starburst galaxy
      (see \S 3.2 for our distinction between AGN and starbursts).

    \item COSMOS J095855.26+022713.7, $i_{\rm AB}^+=22.07$, $z=1.13$,
      $z_{\rm conf}=4$: This narrow emission line spectrum is faint,
      but the \OII~ emission feature has a strong signal above the
      noisy continuum.  We assign this spectrum $z_{\rm conf}=4$
      because there is no other plausible redshift solution for a
      single bright narrow emission line.  The 2D spectrum (not shown)
      also reveals the emission feature in both nodded positions,
      confirming that it is not a noise spike.  This object is a Type
      2 AGN with both $L_{0.5-10~{\rm keV}} > 3 \times 10^{42}~{\rm
      erg~s}^{-1}$ and $-1 \le \log{f_X/f_O} \le 1$.

    \item COSMOS J095806.24+020113.8, $i_{\rm AB}^+=21.26$, $z=0.62$,
      $z_{\rm conf}=4$: We identify this spectrum as a hybrid ``nla''
      object, since it has both narrow emission lines and the
      absorption lines of an early-type galaxy.  H$\beta$ is present
      only in absorption, and while half of the H+K doublet is on a
      masked-out region, the other line is present.

    \item COSMOS J095906.97+021357.8, $i_{\rm AB}^+=21.11$, $z=0.76$,
      $z_{\rm conf}=4$: This spectrum exhibits only absorption lines
      and is classified as an early-type galaxy.  The continuum shape
      and H+K doublet make assigning redshifts to these targets
      straightforward.  This object meets both of the X-ray emission
      criteria of \S 3.2 and is an optically obscured AGN.

    \item COSMOS J095743.85+022239.1, $i_{\rm AB}^+=22.20$, $z=1.02$,
      $z_{\rm conf}=1$: This spectrum is quite noisy.  The single
      narrow line may be \OII, but it is not strong enough above the
      noise to reliably classify.  Because its entire identification
      may be a result of noise, we designate this target as $z_{\rm
      conf}=1$.

\end{enumerate}

\subsection{MMT/Hectospec}

We also obtained ancillary spectroscopic data using the Hectospec
fiber-fed spectrograph \citep{fab05} on the 6.5 m MMT telescope.  The
field of view for Hectospec is a 1 deg diameter circle, and in March
2007 the COSMOS field was observed with two pointings of 3 hours each,
as shown in Figure \ref{fig:field}.  These pointings contained a total
of 134 targets to $i_{\rm AB}^+<23.5$ in 2.5$\arcsec$ fibers.  We
observed with the 270 l/mm grism over a wavelength coverage of
3800-9200\AA, resulting in a resolution of 3\AA.  Because Hectospec is
fiber-fed and the MMT has a brighter sky and generally poorer seeing
than Magellan, MMT/Hectospec cannot reach targets as faint as those
reached by Magellan/IMACS.  Therefore we use Hectospec observations
only as ancillary data on targets which already have IMACS spectra.

The MMT/Hectospec observations were designed primarily to double-check
the redshifts derived from IMACS spectra by adding the bluer
3800-5600\AA~wavelength band.  Figure \ref{fig:linefig} shows the
observed peak wavelength with redshift for the strong broad emission
line in Type 1 AGN.  With IMACS, the limited red wavelength range
means that broad line AGN at $0.4<z<1.9$ and $2.3<z<2.9$ will have
only one observed broad line, as shaded in the figure.  These
potentially ambiguous redshifts can be resolved using the Hectospec
spectra.  Even for targets with non-ambiguous redshifts, the extended
wavelength coverage allows for consistency checks and additional line
measurements.

In Figure \ref{fig:imacsmmt} we show two objects where a
high-confidence redshift could be assigned only after Hectospec
spectra were additionally taken.  The first of these,
095801.45+014832.9, was assigned $z_{\rm conf}=2$ and an incorrect
redshift of 1.3 before the Hectospec data allowed us to correctly
resolve the degeneracy and assign $z_{\rm conf}=4$.  The second
object, 100149.00+024821.8, had been assigned the correct redshift
from its IMACS spectrum but only $z_{\rm conf}=2$, and the Hectospec
data confirmed the otherwise uncertain solution and allowed us to
assign $z_{\rm conf}=4$.  In general, the additional Hectospec spectra
revealed that we were $\sim$75\% accurate in assigning redshifts to
IMACS spectra with degenerate redshift solutions.  (We were better
than the 50\% chance probability because we were occasionally able to
fit to minor features, e.g. FeII/III complexes, weak narrow lines like
\OII~and \NeIV, or general continuum shape.)

We reduced the Hectospec data into 1D linear spectra with sky
subtraction, flux calibration, and cosmic ray rejection using the
publicly available HSRED software (written by R. Cool).  We also used
HSRED to apply an artificial flux calibration to correct the spectral
shape, then flux calibrated the spectra using a mean correction from
objects with both Hectospec and IMACS data.  From the resultant
spectral shape we estimate that this technique has flux errors in the
blue and red ends of the spectra as large as $\sim$20\%.  Since the
analyses are limited to finding redshifts and performing simple line
width measurements, errors of this magnitude are acceptable.

\subsection{SDSS}

We include 76 XMM X-ray sources with spectra previously taken as part
of the Sloan Digital Sky Survey \citep[SDSS,][]{york00}.  With
redshifts already known, these targets were excluded from the main
IMACS survey.  These objects were selected using the publicly
available SDSS Catalog Archive Server (http://cas.sdss.org/astro/),
which uses the SDSS Data Release 6 \citep{sdssdr6}.  Their wavelength
coverage is 3800-9200\AA~ and they have a resolution of 3\AA.  All the
SDSS targets are uniformly bright, with $i_{\rm AB}^+ \lesssim 21$,
and so they would certainly have been successfully observed with IMACS
had their redshifts not been previously known.  Including these SDSS
targets does not introduce any new incompleteness or complication to
the sample.

\section{Spectral Analysis}

Our program as described above is largely motivated as an AGN redshift
survey.  We especially seek Type 1, Type 2, and optically-obscured
(host-dominated AGN), though we also find a small contaminant fraction
of local stars and star-forming galaxies.  We attempt to separate the
population of obscured AGN from star-forming and quiescent galaxies
using X-ray and optical color diagnostics.  A companion paper
\citep{tru08} presents basic line measurements and estimates of black
hole mass for the Type 1 AGN.

\subsection{AGN Classification}

We used three composite spectra from the Sloan Digital Sky Survey
\citep[SDSS,][]{york00} as templates for classifying the objects and
determining their redshifts: a Type 1 (broad emission line) AGN
composite from 2204 sources \citep{q1template}, a Type 2 (narrow
emission line) AGN composite from 291 sources \citep{q2template}, and
a red galaxy composite from 965 sources \citep{etemplate}.  The three
template spectra are shown in Figure \ref{fig:templates}.  We found
that the Type 2 AGN composite gave accurate redshifts for both
star-forming galaxies and AGN with narrow emission lines.  The red
galaxy template was likewise accurate for a variety of absorption line
galaxies, ranging from old stellar systems with strong 4000\AA~breaks
to post-starburst galaxies.  Objects showing a mixture of narrow
emission lines and red galaxy continuum shape and absorption features
were classified as hybrid objects.  We did not use a particular
template for local stars, but stars ranging in temperature from O/B to
M types were easily visually identified.

To calculate redshifts we used a cross-correlation redshift IDL
algorithm in the publicly available {\tt idlspec2d} package written by
D. Schlegel\footnote{publicly available at
http://spectro.princeton.edu/idlspec2d\_install.html}.  This algorithm
used a visually-chosen template to find a best-fit redshift and its
associated 1$\sigma$ error.  As discussed in \S 2.2, all masked-out
regions were ignored in the redshift determination.  Note that the
redshift error returned is probably underestimated for objects with
lines shifted from the rest frame with respect to each other, as is
often the case between high-ionization (e.g. \CIV) and low-ionization
(e.g. \MgII) broad emission lines in Type 1 AGN \citep{lineshifts}.
We manually assigned redshift errors for 6\% (41/677) of objects where
the cross-correlation algorithm failed but we were able to visually
assign a best-fit redshift.

Each object was assigned a redshift confidence according to the
ability of the redshifted template to fit the emission lines,
absorption lines, and continuum of the object spectrum.  If at least
two emission or absorption lines were fit well, or if at least one
line and the minor continuum features were fit unambiguously, the
redshift was considered at least 90\% confident and assigned $z_{\rm
conf}=4$ (64\% of objects).  Objects of $z_{\rm conf}=3$ (8\% of
objects) have only one strong line feature with a continuum or second
less certain feature that make their assigned redshift likely but not
as assured.  We assign $z_{\rm conf}=2$ (8\% of objects) when the
spectrum exhibits only one broad or narrow feature and the calculated
redshift is degenerate with another solution.  Objects of $z_{\rm
conf}=1$ (5\% of objects) are little more than guesses, where a sole
feature is present but has little signal over the noise, such that
even the spectral type classification is uncertain.  (It is notable,
however, that the nod and shuffle observations helped to resolve real
features from noise, since real features must occupy both nodded
positions on the CCD.)  If the signal-to-noise of the object spectrum
was too low for even a guess at the redshift or spectral type, it was
assigned $z_{\rm conf}=0$ (13\% of objects).  We additionally assign
$z_{\rm conf}=-1$ to 13 targets with "broken" slits, severely
contaminated by second order lines or mask cutting errors.  In total,
we were unable to assign redshifts for 15\% of targets.  Duplicate
observations with IMACS and Hectospec indicate that redshift
confidences of 4, 3, 2, and 1 correspond to correct redshift
likelihoods of 97\%, 90\%, 75\%, and 33\%, respectively.  These
duplicate observations are mainly estimated for brighter targets,
however, and so the true likelihoods may be slightly lower.  In total,
we were able to classify 573 spectra with $z_{\rm conf}>0$ and we
designate the 485 spectra with $z_{\rm conf}=3,4$ as
``high-confidence'' objects.  We discuss individual targets spanning
the classification types and confidence levels in \S 2.2 and shown in
Figures \ref{fig:spec1} and \ref{fig:spec2}.

All of the objects observed in the sample are presented in Table
\ref{tbl:agncat}.  The classifications are as follows: ``bl'' for
broad emission line objects (Type 1 AGN), ``bnl'' for objects with
both broad and narrow emission (possibly Type 1.5-1.9 AGN), ``nl'' for
narrow emission line objects (Type 2 AGN and star-forming galaxies),
``a'' for absorption line galaxies, ``nla'' for narrow emission and
absorption line galaxy hybrids, and ``star'' for stars (of varied
spectral type).  We further classify narrow emission and absorption
line spectra as AGN or inactive in \S 3.2 below.  In total, 50\%
(288/573) of the classified targets were designated ``bl'' or ``bnl,''
30\% (171/573) were ``nl'' or ``nla,'' 16\% (92/573) were ``a,'' and
the remaining 4\% (22/573) were stars.  Objects with a question mark
under ``Type'' in Table \ref{tbl:agncat} have too low signal-to-noise
to venture a classification, although many of these objects are
unlikely to be Type 1 or 2 AGN for reasons we discuss in \S 4.  As
mentioned above, objects with $z_{\rm conf}=1$ may be incorrectly
classified.

We summarize the efficiencies, from X-ray sources to targeting to
redshifts, in Table \ref{tbl:efficiency}.

\subsection{AGN, Starbursts, and Quiescent Galaxies}

We use the following X-ray emission diagnostics to classify the 485
high-confidence extragalactic objects as AGN:

\begin{equation}
  L_{0.5-10~{\rm keV}} > 3 \times 10^{42}~{\rm erg~s}^{-1}
\end{equation}
\begin{equation}
  -1 \le \log{f_X/f_O} \le 1,~{\rm where}~\log{f_X/f_O} = \log(f_X) + i_{\rm AB}/2.5 + 5.352
\end{equation}

Each of these criteria have been shown by several authors to reliably
(albeit conservatively) select AGN \citep[e.g., ][]{hor01, ale01,
bau04, bun07}, although it is important to note that bona fide AGN
(e.g., LINERs and other low-luminosity AGN) can be much less X-ray
bright than these criteria.  Equation 1 is derived from the fact that
purely star-forming galaxies in the local universe do not exceed
$L_{0.5-10 \rm keV} \simeq 3 \times 10^{42}$ erg~s$^{-1}$ \citep[e.g.,
][]{fab89, col04}.  The X-ray luminosities of the sources are shown in
Figure \ref{fig:xraylum} along with the X-ray flux limit.  Nearly all
of the Type 1 AGN (marked as crosses) lie above the AGN luminosity
threshold.  Equation 2 is the traditional ``AGN locus'' defined by
\citet{mac88}, shown for the sample in Figure \ref{fig:xraylocus}.
Objects marked with x's have $L_{0.5-10 \rm keV} > 3 \times 10^{42}$
erg~s$^{-1}$, revealing that the two methods heavily overlap, with
94\% (405/432) of the objects that satisfy one of the criteria
additionally meeting both.  Only 53 ``nl'' and ``a'' objects do not
meet either of the X-ray criteria, leaving us with 432 X-ray AGN that
meet either Eq. 1 or Eq. 2 and have high-confidence redshifts.

Using either Equations 1 or 2 selects all of the spectroscopically
identified Type 1 AGN, but it still may exclude some obscured AGN.
The source classification diagnostic diagrams, based on the optical
emission line measurements \citet[][BPT]{bpt81}, are usually quite
effective in classifying narrow emission line spectra as star forming
galaxies or Type 2 AGN, with a sound theoretical basis \citep{kew01}
and use in many surveys \citep[e.g., ][]{kau03, trem04}.  The BPT
diagnostic uses ratios of nebular emission lines (\OIII
$\lambda$5007/\Hb~and \NII $\lambda$6583/\Ha) to distinguish between
thermal emission from star formation and nonthermal AGN emission.
However, there are two limitations to the BPT diagnostic that make it
inapplicable to our sample.  First, most of the object do not have the
appropriate lines in their observed wavelength range: most of the
``nl'' objects are at higher redshift and we are limited by the
spectral range of IMACS.  In addition, accurately measuring the line
ratios requires correcting for absorption in \Ha~and \Hb~from old
stellar populations.  Because the spectra have low resolution and
limited wavelength range, we are unable to accurately fit and account
for stellar absorption.

The color-based diagnostic of \citet{smo08} can be used to further
classify the narrow emission and absorption line spectra which do not
satisfy Equations 1 and 2 but are nonetheless AGN.  This selection
technique is based on the a tight correlation in the local universe
between the emission line flux ratios utilized for the spectroscopic
BPT selection and the galaxies' rest-frame optical colors
\citep{smo08}. The method has been well calibrated on the local
SDSS/NVSS sample in \citet{smo08} and successfully applied to the
radio VLA-COSMOS data \citep{smo08b}.  Following \citet{smo08b} the
rest-frame color for the narrow line AGN was computed by fitting each
galaxy's observed optical-to-NIR SED \citep{cap07}, de-redshifted
using its spectroscopic redshift, with a library of 100,000 model
spectra \citep{bru03}.  \citet{smo08} show that the color diagnostic
is a good statistical measure, but may not be accurate for individual
objects.  So while it further indicates that 17/53 objects are
obscured AGN outside the X-ray criteria, we do not include these
objects as AGN and note only that the sample of 432 high-confidence
X-ray AGN as defined by Equations 1 and 2 probably misses at least
$\sim$17 additional objects.

\section{Completeness}

Our targeting was based solely on the available X-ray data and the
optical flux constraint of $i_{\rm AB}^+<23.5$.  While we were only
able to target 52\% (677/1310) of the available X-ray sources, the
spectra obtained were constrained only by slit placement and so
represent a random subset of the total X-ray population.  Therefore
our completeness limits can be determined from the success rate for
the spectroscopy, which is dependent on optical magnitude, object
type, and redshift.  We characterize and justify the flux limits in \S
4.1 and 4.2, as well as the more detailed redshift completeness in \S
4.3.  Our goal is a purely X-ray and optical flux-limited sample of
AGN, and so in \S 4.4 we account for the spectroscopic incompleteness
to infer the AGN population to $f_{0.5-10 \rm keV} < 1 \times
10^{-15}~{\rm erg~cm^{-2}~s^{-1}}$ and $i_{\rm AB}^+ \le 23$.

\subsection{X-ray Flux Limit}

The first limit on the completeness is the target selection, which is
limited in both X-ray and optical fluxes.  The initial selection
includes all XMM targets with X-ray flux limits of $1 \times
10^{-15}~{\rm erg~cm^{-2}~s^{-1}}$ in 0.5-2 keV or $6 \times
10^{-15}~{\rm erg~cm^{-2}~s^{-1}}$ in the hard 2-10 keV band with
optical counterparts of $i_{\rm AB}^+ \leq 23.5$.  The X-ray flux
limit means that we are complete in X-rays to all AGN with
$L_{0.5-10~{\rm keV}} > 3 \times 10^{42}~{\rm erg~s}^{-1}$ (a classic
AGN definition discussed in \S 3.2) at $z \lesssim 1$.

\subsection{Optical Flux Limit}

Our initial magnitude cut was $i_{\rm AB}^+ \leq 23.5$, but this was
designed to be more ambitious than the capabilities of Magellan/IMACS
in 5-hour exposures.  In Figure \ref{fig:sni} we show the spectral
signal-to-noise (S/N) with optical $i_{\rm AB}^+$ magnitude for the
targets from all IMACS exposures.  The S/N was calculated by
empirically measuring the noise in the central 6600-8200\AA~region of
each spectrum.  The S/N generally correlates with the optical
brightness, with some scatter attributable to varied conditions over
the three years of observations.  The outliers with high S/N and faint
magnitude are all emission line sources where a strong emission line
lies in the spectrum but outside the observed $i_{\rm AB}^+$ filter
range.  The low-S/N and bright magnitude objects of the lower left may
be highly variable sources or targets with photometry contaminated by
blending or nearby bright stars.  The increasing number of
unclassified targets (filled green circles) in Figure \ref{fig:sni}
shows that we do not identify all objects to $i_{\rm AB}^+ \leq 23.5$.

In Figure \ref{fig:typesi} we show the completeness with $i_{\rm
AB}^+$ magnitude for the various classifications.  We assume that the
identified fractions have Poisson counting errors from the number of
the given type and the total number of targets in each magnitude bin.
The survey completeness to all targets remains at $\sim$90\% to
$i_{\rm AB}^+<22$.  The identification fractions of emission line
targets remains nearly flat a magnitude deeper than the absorption
line galaxies, although the fractions of ``bl'' and ``nl'' objects
decrease slightly from $22<i_{\rm AB}^+<23$, within the noise.

The completeness is not uniform for all types of objects: the fraction
of identified broad and narrow emission line targets remains
statistically constant until $i_{\rm AB}^+ \sim 22.5$, while the
fraction of absorption line targets appears complete only to $i_{\rm
AB}^+ \sim 22$.  Both narrow and broad emission lines generally
exhibit two or more times the signal of their continuum, allowing for
identification even when the objects' broad-band magnitude and average
S/N are low.  Since different emission lines vary in strength, this
also suggests that the identification of ``bl'' and ``nl'' may suffer
from a redshift dependence (for instance, some redshifts may have only
weak emission lines in their wavelength range, while others include
several strong lines).

\subsection{Redshift-Dependent Completeness}

The strongest redshift dependence in the spectra come from targets
with only one strong emission line in their observed IMACS spectra.
The ``a'' type objects are well-populated with absorption lines and
have redshift-independent classifications, but emission line spectra
may have only one line in the observed 5600-9200 \AA~window (see
Figure \ref{fig:linefig}).  The presence of only one emission line
causes two problems: the redshift solution will be degenerate, and the
line may be confused with noise if it is either narrow or broad but
weak.  The first problem means that we can only assign $z_{\rm
conf}=2$ and we may also assign the wrong redshift (bright targets are
an exception, since a redshift can be assigned based on strong
continuum features).  The second problem means that we might
completely miss the AGN designation and assign it a ``?''
classification.  At faint S/N levels, the pattern of two emission
lines is much easier to identify over the noise, and so the lower
completeness to single-line objects may mean both lower redshift
confidence and a lower identification threshold.

We used Monte Carlo simulations to test the redshift and magnitude
dependence of the survey's completeness for emission line spectra.  We
assume that the SDSS Type 1 composite spectrum \citep{q1template} and
Type 2 composite spectrum \citep{q2template} each have infinite
signal-to-noise, and degrade these spectra with Gaussian-distributed
random noise to represent broad and narrow emission line spectra of
varying $i_{\rm AB}^+$ magnitude.  For each bin in magnitude, we
calculate the median signal-to-noise of the observed spectra at that
brightness, measured at both $\lambda<8000$\AA~and in the noisier
region with sky lines at $\lambda>8000$\AA~(our spectra typically have
S/N about 16\% worse at $\lambda>8000$\AA).  Each artificial
noise-added spectrum was then redshifted over several values and
realized in the IMACS wavelength range (5600\AA-9200\AA).  We then
used the same {\tt idlspec2d} redshift algorithm used on the data
described in \S 3.1 to determine whether or not we would be able to
assign the correct redshift with high-confidence ($z_{\rm conf}=3,4$)
for these artificial redshifted spectra (a redshift could not be
determined if the emission lines were smeared out or if the spectrum
could not be distinguished from noise or a different line at another
redshift).  We used 20 realizations for each redshift and
signal-to-noise bin.  The fraction of artificial spectra with
determined redshifts at a given redshift and signal-to-noise, with
different seeds of randomly-added noise, forms an estimate of the
completeness.

We found that the simulated completeness for narrow emission line
spectra was 90\% complete to $i_{\rm AB}^+ \sim 23$ ($S/N \approx
1.76$ per pixel) for $z \le 1.3$, with strong unambiguous lines (e.g.,
\Ha$\lambda$6563, \Hb$\lambda$4861, \OIII$\lambda$5007,
\OII$\lambda$3727).  This is a magnitude fainter than the level of the
average redshift completeness of the survey.  At $0.9 \le z \le 1.4$,
\OII$\lambda$3727 is the only strong line, but it is bright enough
that the redshift solution remains unambiguous even to $i_{\rm AB}^+ <
23$.  At $z>1.4$ the \OII~ line shifts completely out of the
wavelength range and no good emission lines remain.  The additional
blue Hectospec coverage is also useless at $z>1.4$, since \OII~remains
redward of the upper 9200\AA~wavelength limit.  We cannot identify
narrow emission line (``nl'') spectra at $z>1.4$.

The Type 1 AGN completeness has a more complex redshift dependence.
As shown in Figure \ref{fig:linefig}, in the redshift ranges $0.4
\lesssim z \lesssim 1.9$ and $2.3 \lesssim z \lesssim 2.9$ only one
line is present and the redshift solution may be degenerate.  This is
ameliorated by the ancillary MMT/Hectospec spectra which have broader
wavelength coverage.  Examples of two objects with only one emission
line in their IMACS spectra, but two emission lines in their Hectospec
spectra, are shown in Figure \ref{fig:imacsmmt} with accompanying
discussion in \S 3.3.  Only 36\% (104/288) of the broad emission
(``bl'') spectra benefit from MMT/Hectospec coverage.  We add this
MMT/Hectospec corroboration to the unidentified targets in the
simulations, and estimate the redshift completeness as shown in Figure
\ref{fig:q1complete}.  We have lower redshift completeness in the
redshift ranges $0.5 \lesssim z \lesssim 1.5$, and $2.3 \lesssim z
\lesssim 2.6$, where only one line is present ($\Hb$, $\MgII$, or
$\CIII$) and although we can reliably classify as a broad line AGN
(``bl'') it is difficult to distinguish between the two redshift
ranges.  Without the degeneracies between redshift, the redshifts
would be $>90\%$ complete to $S/N \approx 1.75$ (per pixel) or $i_{\rm
AB}^+ \sim 23$).

We do not test redshift dependence in identifying absorption line
(``a'') spectral types because these spectra are well-populated with
absorption lines.  At $z>1.3$ the 4000\AA~break leaves the wavelength
range, but otherwise each absorption line galaxy has the same aptitude
for $z_{\rm conf} \ge 3$ classification at $z<1.3$.  However, because
the absorption line (``a'') spectra lack features which are of higher
signal than their continua, we cannot identify them to the same low
S/N levels as emission line spectra.  So the incompleteness to
absorption line (``a'') spectra at $z<1.3$ with $22 < i_{\rm AB}^+ <
23$ is not redshift dependent.  Because we have high completeness to
broad line AGN (``bl'') and narrow emission line spectra (``nl'') at
this magnitude (excepting the redshift ranges described above), most
of the unidentified targets at $22 < i_{\rm AB}^+ < 23$ are probably
absorption line galaxies.

In summary, the sample has the following incompleteness outside of the
flux limits:

\begin{enumerate}
  \item Type 1 AGN of $22<i_{\rm AB}^+<23$ at $z \sim 0.8$, $z \sim
    1.3$, and $z \sim 2.4$ (completeness in these regions shown in
    Figure \ref{fig:q1complete}).
  \item Type 2 AGN of all magnitudes at $z>1.4$
  \item Absorption line galaxies of $22<i_{\rm AB}^+<23$ at $z<1.3$
    (from \S 4.2), and absorption line galaxies of all magnitudes at
    $z>1.3$
\end{enumerate}

We show the redshift distribution of all AGN (meeting one of the X-ray
criteria in \S 3.2) in Figure \ref{fig:zhist}.  The uncorrected
redshift distribution is shown by the square symbols.  We next attempt
to describe the complete $i_{\rm AB}^+<23$ flux-limited sample,
correcting for the incompleteness of the four points above.

\subsection{Characterizing the Low-Confidence Targets}

We can only assign high-confidence ($z_{\rm conf}=3,4$) redshifts for
72\% (485/677) of the targets, leaving 88 spectra with low-confidence
($z_{\rm conf}=1,2$ redshifts and 104 targets of unknown spectral type
($z_{\rm conf}=-1,0$).  We characterize these 192 low-confidence and
unclassified spectra using the photometric classifications and
redshifts of \citet{sal08}, which take advantage of the extensive
photometry of COSMOS \citep{cap07}.  The photometric redshift
algorithm finds a best-fit redshift and classification by matching a
set of 30 templates to the IR (IRAC), optical (Subaru), and UV (GALEX)
photometric data of each object.  The templates are described in full
detail in \citet{sal08} and are available upon request\footnote{Mara
Salvato, ms@astro.caltech.edu}.  The photometric redshift technique
was calibrated upon the spectroscopic redshifts we present for the 485
spectra of high redshift confidence, and has a precision of
$\sigma_{\Delta z} / (1+z) < 0.02$ with $<5\%$ of targets as
significant outliers at $z<4.5$.

The photometric redshift templates rely on multiwavelength fitting
from IR to UV wavelengths, and so the photometric classifications can
separate AGN-dominated (which we designated ``unobscured'') and
host-dominated (which we designate ``obscured'') AGN types.  However,
the photometric classifications do not distinguish well between our
absorption line spectra (``a'' types) and narrow emission line spectra
(``nl'' types), although they can separate unobscured broad line AGN
(``bl'' types) from obscured AGN (``a'' and ``nl'' types).  We must
assume population fractions of absorption line and narrow emission
line spectra from the photometrically classified obscured objects
using the known fractions from the high-confidence spectroscopy.  In
Figure \ref{fig:typesi}, the fraction of narrow emission (``nl'')
spectra does not decrease appreciably to $i_{\rm AB}^+ \sim 22.5$ and
almost all of the unknown objects can be assumed to be absorption line
(``a'') types.  We also know from \S 4.3 above that we are incomplete
to ``nl'' spectra at $z>1.4$, and the spectroscopically unclassified
targets at $z>1.4$ probably follow the $\sim$2:1 ratio of narrow
emission (``nl'') to absorption (``a'') types we find at lower
redshifts in \S 3.1.  So we assume that all photometrically classified
unobscured AGN correspond to our broad emission (``bl'') type, and
assume fractions of absorption (``a'') and narrow emission spectra
(``nl'') as follows: (1) for $z<1.4$, all are ``a'' types, and (2) for
$z>1.4$, 2/3 are ``nl'' types and the remainder are ``a'' types.

Figure \ref{fig:badphotz} shows the photometric redshift distribution
for the 192 low confidence and unclassified spectra.  Most of the
photometric redshifts fall into one of the three incompleteness
categories shown above in \S 4.3.  We will use this redshift
distribution to characterize the demographics of the complete
flux-limited sample in \S 5.

We can also use the absolute magnitude distribution of the targets
with secure spectroscopic redshifts in Figure \ref{fig:absmag} to make
a qualitative assessment of the unidentified targets.  The dashed
lines mark $i_{\rm AB}^+=22$ and $i_{\rm AB}^+=23$.  We will assume
that the absolute magnitude distribution for narrow emission (``nl'')
and absorption (``a'') objects, which peaks at $M_i \sim -22$, does
not change with redshift.  Objects of this absolute magnitude
distribution should be detected to $z \sim 2$, but there are no narrow
emission (``nl'') spectra detected at $z>1.4$, and no absorption
(``a'') spectra detected at $z>1.3$.  So many of the unidentified
targets are probably $z>1.4$ ``nl'' and $z>1.3$ ``a'' type objects.
The bright tail of the $M_i$ distribution for ``a'' and ``nl'' types
at $z>1.3$/$z>1.4$ also has $i_{\rm AB}^+<22$, and so these missing
$z>1.3$/$z>1.4$ targets may account for the unidentified targets at
$i_{\rm AB}^+<22$.  In addition, most obscured AGN have $M_i \sim
-22$, which lies within $22<i_{\rm AB}^+<23$ at $z>1.3$/$z>1.4$,
suggesting that these objects may be most of the unidentified
$22<i_{\rm AB}^+<23$ targets.  This qualitative assessment confirms
the characterization of the unknown spectral types using photometric
redshifts.

\section{Discussion}

\subsection{Demographics}

Figure \ref{fig:xraylum} indicates that the X-ray flux limit includes
all $L_{0.5-10 \rm keV} > 3 \times 10^{42}$ erg s$^{-1}$ AGN at $z<1$.
This means we are nearly complete to all X-ray AGN (as defined in \S
3.2) at $z<1$, since almost all objects that meet one of the X-ray
criteria also meet both.  We can additionally see in Figure
\ref{fig:absmag} that we observe all but the faint tail of the $M_i$
distributions of obscured and unobscured AGN types to $z<1$, so long
as we use the simple corrections of \S 4.4 to characterize the sample
to $i_{\rm AB}^+<23$.  This allows us to characterize the complete
$z<1$ X-ray AGN population.

In Figure \ref{fig:zhist} we show the number of each AGN type with
redshift.  This includes only the 432 high-confidence X-ray AGN as
defined by the X-ray criteria.  We find raw fractions of broad
emission line (56\%), narrow emission line (32\%), and absorption line
(12\%) over all redshifts, which roughly agree with other wide-area
X-ray surveys \citep{fio03, sil05, eck06, tru07}.  To characterize the
complete $i_{\rm AB}^+<23$ and $f_{0.5-10 \rm keV} > 1 \times
10^{-15}$ erg s$^{-1}$ cm$^{-2}$ sample, we include the 106 $i_{\rm
AB}^+<23$ targets with bad spectroscopy and photometric redshifts that
satisfy the X-ray AGN criteria.  The corrected fractions of
$i_{AB}^+<23$ targets at all redshifts include 57\% broad emission
line, 25\% narrow emission line, and 18\% absorption line AGN.

\subsection{Obscured to Unobscured AGN Ratio}

The ratio of obscured to unobscured AGN can help determine the
properties of the obscuration which hides nuclear activity.  In the
simplest unification models \citep{ant93,urr95}, obscuration depends
only on the orientation and should remain independent of luminosity
and redshift.  However, since we know that galaxies at higher
redshifts have more dust than local galaxies, then one might expect
the ratio of obscured to unobscured AGN to depend on redshift if AGN
host galaxy dust plays a role in obscuration \citep[e.g.,][]{bal06}.
And if the obscuring dust (or its sublimation radius) is blown out
further by more luminous accretion disks \citep{law82, law91, sim05},
then one might expect the ratio to decrease with increasing
luminosity.  Some models of the X-ray background prefer ratios which
suit these physical descriptions, predicting an increasing ratio of
obscured to unobscured with increasing redshift and decreasing
luminosity \citep{bal06, trei06}.  Deep X-ray observations confirm
that the ratio depends on luminosity \citep{ste04, bar05, trei08}.
Some observations additionally suggest redshift evolution
\citep{laf05, trei06, has08}, but other authors claim that redshift
evolution is neither necessary in the models nor significant in the
observations \citep{ued03, aky06, gil07}.

We derive the obscured to unobscured AGN ratio with redshift in Figure
\ref{fig:obscurfrac}.  Here ``obscured AGN'' refers to both narrow
emission line (``nl'') and absorption line (``a'') AGN meeting the
X-ray criteria of \S 3.2, and ``unobscured AGN'' includes all
broad-line (``bl'') AGN.  To the limit of the survey at $z<1$, our
average ratio is 3:1 obscured to unobscured AGN.  We additionally
separate the AGN into X-ray luminous and X-ray faint (in relation to
the median X-ray luminosity, $L_{\rm med}=1.32 \times 10^{44}$ cgs) in
the bottom panel of Figure \ref{fig:obscurfrac}.  The ratio of
obscured to unobscured X-ray faint AGN appears to be much higher than
the ratio for X-ray bright AGN, and additionally seems to increase
with redshift.

We test the ratio for dependence on redshift and luminosity using
logistic regression a useful method for determining how
classification depends upon a set of variables.  It is commonly used
in biostatistical applications, where one expects a binary response
(for instance, a patient might live or die) based on a set of
variables.  Logistic regression considers each data element
independently, and is therefore more effective than significance tests
which bin the data.  An excellent review of logistic regression is
found in \citet{fox97}.  We use the method here to learn if the
likelihood for an AGN to be classified as obscured or unobscured (a
binary response) depends on observed X-ray luminosity and/or redshift.
Logistic regression solves for the ``logit'' (the natural logarithm of
the odds ratio) in terms of the variables as follows:

\begin{equation}
  \ln{\frac{Pr(G=1|L,z)}{Pr(G=2|L,z)} = \beta_0 + \beta_1\ln(1+z) + \beta_2\ln(L_X/10^{42})}
\end{equation}

Here $G=1$ means an AGN is classified obscured, and $G=2$ means an AGN
is classified unobscured.  We use $\log(1+z)$ and $\log(L_X/10^{42})$
as the dependent variables instead of $z$ and $L_X$ for numerical
stability.  Then the logit, as the logarithm of the ratio of the
probabilities, is just the logarithm of the obscured to unobscured
ratio.  We solve for the coefficients using the Newton-Raphson method
and estimate errors by bootstrapping, calculating the standard
deviation on the coefficients with 1000 random subsets of the true
data.  We find the coefficients to be

\begin{equation}
  \beta_0 = 1.73 \pm 0.67;~ \beta_1 = 3.83 \pm 1.57;~ \beta_2 = -0.69 \pm 0.19
\end{equation}

In other words, the obscured/unobscured AGN ratio increases with
redshift at 2.4$\sigma$ significance and decreases with observed X-ray 
luminosity at 3.6$\sigma$ significance.  We can write the power-law 
equation of the expected ratio for a given luminosity and redshift as:

\begin{equation}
  \frac{Pr({\rm obscured})}{Pr({\rm unobscured})} \propto 5.6(1+z)^{3.8}(L_X/10^{42})^{-0.7}
\end{equation}

The curves from this logistic regression model are shown in Figure
\ref{fig:obscurfrac}.  In the bottom panel, the red line shows the
power-law relation (Equation 5) computed using $L_X = L_{\rm low}$,
where $L_{\rm low}$ is the median luminosity from only those AGN with
$L_{0.5-10~ \rm keV}<L_{\rm med}$.  Similarly the blue line represents
the relation for higher-luminosity AGN of $L_{0.5-10~ \rm keV}>L_{\rm
med}$.  Note that the data as binned in Figure \ref{fig:obscurfrac}
shows less signal than the independent data used in the logistic
regression fit, and so the fit of the power-laws shown should not be
judged by the basis of their by-eye match to the binned data.  It is
worth noting, however, that high-luminosity sources seem to evolve
much more weakly with redshift.  This is a natural consequence of the
power-law nature of Equation 5: when $L_X$ is large, the
obscured/unobscured ratio becomes small, and so it appears only to
weakly evolve with redshift on a linear scale.  Our dependence of
obscuration on redshift and luminosity are both consistent with recent
work by both \citet{trei06} and \citet{has08}, with the obscured
fraction about 4 times higher at low luminosity than at high
luminosity and about 2 times higher at $z \sim 1$ than at $z \sim 0$.

The trend with observed X-ray luminosity can be explained in several
ways.  It may be that obscured AGN are simply more absorbed in the
X-rays, such that their intrinsic X-ray luminosities are significantly
higher than their observed.  Then the apparent lack of obscured AGN at
higher X-ray luminosities might be only an observed effect, and not an
intrinsic physical effect.  But if the intrinsic and observed X-ray
luminosities are not significantly different in these obscured AGN,
then the luminosity dependence indicates that more luminous AGN have
less obscuring material.  The luminosity may decrease the opening
angle of obscuration by causing dust sublimation to occur at larger
radii.

The presence of more obscured AGN at $z \sim 1$ is not likely to
indicate physical evolution in AGN, since AGN at similar luminosities
at $z \sim 1$ and $z \sim 0$ are not observed to have different
physical properties like black hole mass and accretion rate
\citep{kel08} or spectral energy distribution \citep{vig03, ric06,
hop07}.  However, galaxies at $z \sim 1$ show significantly more star
formation, gas, and dust than galaxies at $z \sim 0$, and so the
increase of obscuration with redshift may be explained by host
gas/dust obscuration of the AGN central engine.  Indeed, models by
\citet{bal08} show that star formation can effectively obscure AGN
while producing both the observed luminosity and redshift dependence
of the obscured/unobscured ratio.  \citet{bal08} additionally show
that starburst-driven obscuration should be easily distinguished from
AGN-heated dust by future Herschel 100 $\mu$m surveys.

It is important to note that our definition of ``obscured'' includes
only moderately X-ray obscured AGN.  We are not sensitive to
Compton-thick and other heavily X-ray obscured AGN, and so may be
significantly underestimating the obscured AGN population
\citep{dad07, fio08}.  Logistic regression reveals statistically
significant evidence for redshift evolution and dependence on X-ray
luminosity in the optically obscured/unobscured ratio, but mid-IR
surveys may reveal different dependencies by including heavily
obscured AGN missed in X-rays.

\section{Conclusions and Future Projects}

We present optical spectroscopy for 677 X-ray targets from COSMOS,
with spectra from Magellan/IMACS, MMT/Hectospec, and archival SDSS
data.  The spectroscopy is uniformly complete to $i_{AB}^+<22$.  By
using photometric redshifts for the bad spectra, we additionally
characterize the sample to $i_{AB}^+<23$, and we show that this
optical limit, along with our X-ray flux limit, allows us to
characterize a solely volume-limited sample of all (obscured and
unobscured) X-ray AGN at $z<1$.  We provide evidence that at $z<1$,
the ratio of obscured to unobscured AGN increases with redshift and
decreases with luminosity, where the redshift dependence is of
moderate statistical significance (2.4$\sigma$) and the luminosity
dependence is of higher statistical significance (3.6$\sigma$).

Despite such leverage in the sample presented here, the observations
of AGN in COSMOS are by no means complete.  We were only able to
target 52\% of the available $i_{AB}^+<23.5$ XMM targets, and we hope
to include the remainder of targets in future spectroscopic
observations.  Some of these targets were observed on Magellan/IMACS
and MMT/Hectospec in March 2008, and many of the other XMM targets
without spectra will be observed with VLT/VIMOS (at 5600-9400 \AA) as
part of the zCOSMOS galaxy redshift survey \citep{lil08}.  The zCOSMOS
survey will additionally target $i_{AB}^+>23.5$ XMM targets which are
too faint for Magellan/IMACS.  It is also possible to study fainter
X-ray sources, since the 0.5-2 keV {\it Chandra} observations in
COSMOS go to $2 \times 10^{-16}~{\rm erg~cm^{-2}~s^{-1}}$ in the
central 0.8 deg$^2$, five times fainter than the XMM observations used
here.  Optical identification of these sources are still ongoing, but
the Chandra data are expected to reveal twice as many X-ray targets as
the XMM-selected targets presented here.  We will additionally use the
previously observed spectra of radio and infrared selected AGN
candidates to study Compton-thick and other X-ray faint AGN,

Future work will also use the bolometric studies made possible by the
deep multiwavelength coverage of COSMOS.  We plan to further study the
evolution of obscuration with more fundamental physical quantities
like bolometric luminosity.  A companion paper \citep{tru08} presents
virial black hole mass estimates for the Type 1 AGN presented here and
suggests that it is difficult to form a broad line region below a
critical accretion rate, as suggested previously by \citet{nic00} and
\citet{kol06}.  This concept, combined with the luminosity evolution
of the obscuration presented here, suggests that models of the AGN
central engine must include a prescription where the amount of
obscuring material decreases with increasing luminosity, accretion
rate, or both.

\acknowledgements

We thank the anonymous referee for thorough and helpful comments.  We
additionally thank Alan Dressler and the IMACS team for creating an
excellent instrument, as well as telescope operators Hernan Nu\~{n}ez,
Felipe Sanchez, Geraldo Valladares, Sergio Vera, Hugo Rivera, and the
Las Campanas Observatory staff for support while observing.  We thank
Mike Westover, Alison Coil, Scott Burles, and the Carnegie queue mode
scientists for help with some of the IMACS observations.  We thank
Daniel Fabricant for designing the excellent instrument Hectospec, and
thank MMT operators and Hectospec queue mode scientists for their help
during observations.  We also thank the COSMOS team for their work in
creating the catalogs used for selecting our targets.  We thank
Patrick Shopbell, Anastasia Alexov, and the NASA IPAC/IRSA team for
assisting in uploading the data and catalogs to public IRSA archives.
Observations reported here were obtained at the Magellan Telescopes,
which is operated by the Carnegie Observatories, as well as the MMT
Observatory, a joint facility of the University of Arizona and the
Smithsonian Institution.  The HST COSMOS Treasury program was
supported through NASA grant HST-GO-09822.  JRT acknowledges support
from an ARCS foundation fellowship.

\clearpage
\begin{center}

\begin{deluxetable}{rccc|cc}
\tablecolumns{6}
\tablewidth{0pc}
\tablecaption{COSMOS Observation Log of the IMACS Spectroscopic Observations
\label{tbl:imacsnights}}
\tablehead{
  \colhead{IMACS} &
  \multicolumn{2}{c}{Center (J2000)} & 
  \colhead{Observation} & 
  \colhead{Exposure} &
  \colhead{Number of} \\
  \colhead{Field} &
  \colhead{RA (hh:mm:ss)} &
  \colhead{Dec (dd:mm:ss)} &
  \colhead{Year} &
  \colhead{(hours)} &
  \colhead{Spectra Extracted} }
\startdata
 1 & 09:58:24 & 02:42:34 & 2006 & 4.52 & 33 \\
 2 & 09:59:48 & 02:42:30 & 2007 & 4.00 & 46 \\
 3 & 10:01:06 & 02:42:38 & 2006,2007\tablenotemark{a} & 6.00 & 75 \\
 4 & 10:02:33 & 02:42:34 & 2006 & 5.33 & 37 \\
 5 & 09:58:26 & 02:21:25 & 2006,2007\tablenotemark{a} & 6.00 & 56 \\
 6 & 09:59:47 & 02:21:25 & 2005 & 6.90 & 43 \\
 7 & 10:01:10 & 02:21:25 & 2005 & 6.03 & 48 \\
 8 & 10:02:36 & 02:21:29 & 2006 & 5.10 & 42 \\
 9 & 09:58:25 & 02:00:13 & 2006 & 5.03 & 27 \\
10 & 09:59:47 & 02:00:17 & 2005 & 6.64 & 35 \\
11 & 10:01:10 & 02:00:17 & 2005 & 4.67 & 39 \\
12 & 10:02:37 & 02:02:05 & 2005 & 4.77 & 37 \\
13 & 09:58:24 & 01:39:08 & 2006 & 2.67 &  7 \\
14 & 09:59:47 & 01:39:08 & 2006 & 5.33 & 23 \\
15 & 10:01:10 & 01:39:08 & 2005 & 3.63 & 25 \\
16 & 10:02:33 & 01:39:08 & 2005 & 3.93 & 28 \\
\enddata
\tablenotetext{a}{Fields 3 and 5 were observed for one hour in 2006
and five hours in 2007, for six total hours of exposure.}
\end{deluxetable}

\clearpage
\begin{landscape}
\begin{deluxetable}{l|rr|rrr|crrc}
\tablecolumns{10}
\tablewidth{600pt}
\tablecaption{COSMOS XMM Optical Spectroscopy Catalog
\label{tbl:agncat}}
\tablehead{
  \colhead{Object Name} & 
  \colhead{RA (J2000)} & 
  \colhead{Dec (J2000)} & 
  \colhead{$i_{\rm CFHT}^+$} & 
  \colhead{S/N} & 
  \colhead{$t_{\rm exp}$} & 
  \colhead{Type} & 
  \colhead{z} & 
  \colhead{$\sigma_z$} & 
  \colhead{$z_{\rm conf}$\tablenotemark{b}} \\
  \colhead{} & 
  \multicolumn{2}{c}{degrees (J2000)\tablenotemark{a}} &
  \colhead{AB mag} & 
  \colhead{} & 
  \colhead{sec} & 
  \colhead{} & 
  \colhead{} & 
  \colhead{} &
  \colhead{} }
\startdata
  SDSS J095728.34+022542.2 &  149.3680700 &    2.4283800 & 19.64 &   7.00 &     0 &    bl &  1.5356 &  0.0015 &  4 \cr
COSMOS J095740.78+020207.9 &  149.4199229 &    2.0355304 & 21.55 &  17.92 & 19200 &    bl &  1.4800 &  0.0028 &  4 \cr
  SDSS J095743.33+024823.8 &  149.4305400 &    2.8066200 & 20.43 &   3.37 &     0 &    bl &  1.3588 &  0.0020 &  4 \cr
COSMOS J095743.85+022239.1 &  149.4327000 &    2.3775230 & 23.40 &   1.33 & 18000 &    nl &  1.0192 &  0.0002 &  1 \cr
COSMOS J095743.95+015825.6 &  149.4331452 &    1.9737751 & 21.91 &   1.54 & 19200 &     a &  0.4856 &  0.0030 &  1 \cr
COSMOS J095746.71+020711.8 &  149.4446179 &    2.1199407 & 20.78 &  11.37 & 19200 &    bl &  0.9855 &  0.0002 &  4 \cr
COSMOS J095749.02+015310.1 &  149.4542638 &    1.8861407 & 20.36 &  13.68 & 19200 &   nla &  0.3187 &  0.0002 &  4 \cr
COSMOS J095751.08+022124.6 &  149.4628491 &    2.3568402 & 20.73 &   3.91 & 18000 &   bnl &  1.8446 &  0.0001 &  4 \cr
COSMOS J095752.17+015120.1 &  149.4673623 &    1.8555716 & 21.08 &   7.31 & 19200 &    bl &  4.1744 &  0.0005 &  4 \cr
COSMOS J095753.44+024114.2 &  149.4726733 &    2.6872864 & 22.18 &   0.95 & 11160 &    bl &  2.3100 & -1.0000 &  1 \cr
COSMOS J095753.49+024736.1 &  149.4728835 &    2.7933716 & 21.96 &   4.76 & 11160 &    bl &  3.6095 &  0.0128 &  4 \cr
  SDSS J095754.11+025508.4 &  149.4754500 &    2.9189900 & 19.45 &   6.09 &     0 &    bl &  1.5688 &  0.0022 &  4 \cr
  SDSS J095754.70+023832.9 &  149.4779200 &    2.6424700 & 19.35 &   8.04 &     0 &    bl &  1.6004 &  0.0015 &  4 \cr
  SDSS J095755.08+024806.6 &  149.4795000 &    2.8018400 & 19.41 &   8.66 &     0 &    bl &  1.1108 &  0.0017 &  4 \cr
COSMOS J095755.48+022401.1 &  149.4811514 &    2.4003076 & 21.26 &  19.89 & 18000 &    bl &  3.1033 &  0.0003 &  4 \cr
COSMOS J095756.77+024840.9 &  149.4865392 &    2.8113728 & 20.81 &  11.86 & 11160 &    bl &  1.6133 &  0.0098 &  3 \cr
COSMOS J095757.50+023920.1 &  149.4895683 &    2.6555795 & 20.30 &  11.17 & 11160 &    nl &  0.4674 &  0.0002 &  2 \cr
  SDSS J095759.50+020436.1 &  149.4979100 &    2.0766900 & 18.98 &  14.57 &     0 &    bl &  2.0302 &  0.0016 &  4 \cr
COSMOS J095800.41+022452.5 &  149.5017000 &    2.4145710 & 22.57 &   3.53 & 18000 &   bnl &  1.4055 &  0.0001 &  4 \cr
COSMOS J095801.34+024327.9 &  149.5055777 &    2.7244216 & 20.66 &   9.67 & 11160 &   nla &  0.3950 &  0.0010 &  1 \cr
COSMOS J095801.45+014832.9 &  149.5060326 &    1.8091427 & 21.96 &   1.79 &  9600 &    bl &  2.3995 &  0.0002 &  4 \cr
COSMOS J095801.61+020428.9 &  149.5067217 &    2.0746879 & 22.18 &   5.46 & 19200 &    bl &  1.2260 & -1.0000 &  1 \cr
COSMOS J095801.78+023726.2 &  149.5074058 &    2.6239318 & 17.79 &  52.98 & 11160 &  star &  0.0000 &  0.0000 &  4 \cr
COSMOS J095802.10+021541.0 &  149.5087524 &    2.2613900 & 21.01 &   3.75 &  3600 &     a &  0.9431 &  0.0050 &  3 \cr
\enddata
\tablenotetext{a}{The RA and Dec refer to the optical counterpart of
the X-ray source, which is where the slit was centered.}
\tablenotetext{b}{From empirical measurements, the redshift confidence
was found to correspond to correct redshift likelihoods of 97\%, 90\%,
75\%, and 33\% for $z_{\rm conf}=4,3,2,1$, respectively.  The redshift
confidences are fully explained in \S 3.1.}
\end{deluxetable}
\clearpage
\end{landscape}

\begin{deluxetable}{lrrrr}
\tablecolumns{5}
\tablewidth{0pc}
\tablecaption{Targeting and Redshift Yields
\label{tbl:efficiency}}
\tablehead{
  \colhead{} &
  \colhead{16-Field} & 
  \multicolumn{3}{c}{Per Field} \\
  \colhead{X-ray Sources} &
  \colhead{Total} &
  \colhead{Minimum} &
  \colhead{Maximum} &
  \colhead{Median} }
\startdata
All Sources & 1640 & 68 & 145 & 105 \\
$i_{\rm AB}^+<23.5$ & 1310 & 55 & 110 & 86 \\
Targeted & 677 & 9 & 74 & 38 \\
Classified ($z_{\rm conf}>0$) & 573 & 9 & 63 & 30 \\
$z_{\rm conf}=3,4$ Redshifts & 485 & 8 & 53 & 26 \\
$z_{\rm conf}=3,4$ with Hectospec & 117 & 1 & 27 & 6 \\
$z_{\rm conf}=3,4$ with SDSS & 76 & 2 & 12 & 4 \\
\enddata
\end{deluxetable}

\clearpage

\begin{figure}
\plotone{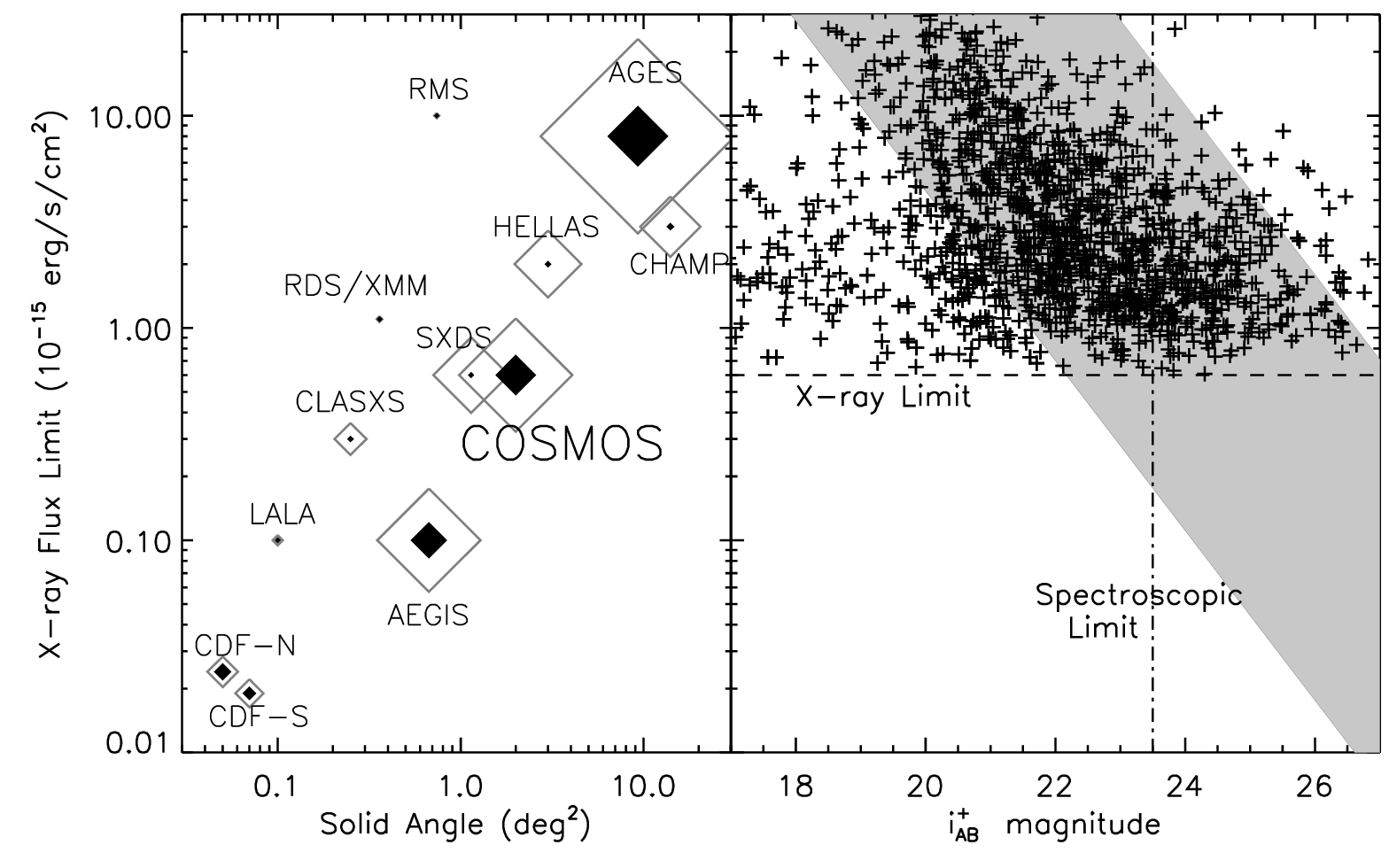}
\figcaption{The X-ray depth and survey size of various deep X-ray AGN
surveys, along with the X-ray and optical flux for targets in COSMOS.
At left, symbol sizes indicate each survey's number of X-ray point
sources: open indicate all sources, and filled indicate those with
optical spectroscopy.  References for the surveys are as follows:
AEGIS \citep{dav07}, AGES \citep{bra06}, CDF-N \citep{ale03,bar03},
CDF-S \citep{luo08}, CHAMP \citep{kim04,gre04}, CLASXS \citep{yang04},
HELLAS2XMM \citep{fio03,coc07}, LALA \citep{wang04}, RDS/XMM
\citep{leh01}, RMS \citep{has05}, and SXDS \citep{ued08}.  At right,
the crosses represent all XMM point sources from \citet{bru08} and the
gray shaded area represents the ``AGN locus'' of
$-1<$~log$(f_X/f_O)<1$ \citep{mac88}.  The COSMOS Chandra data (not
presented here) go four times deeper in the central 0.8 deg$^2$,
doubling the number of COSMOS point sources.
\label{fig:surveys}}
\end{figure}

\begin{figure}
\plotone{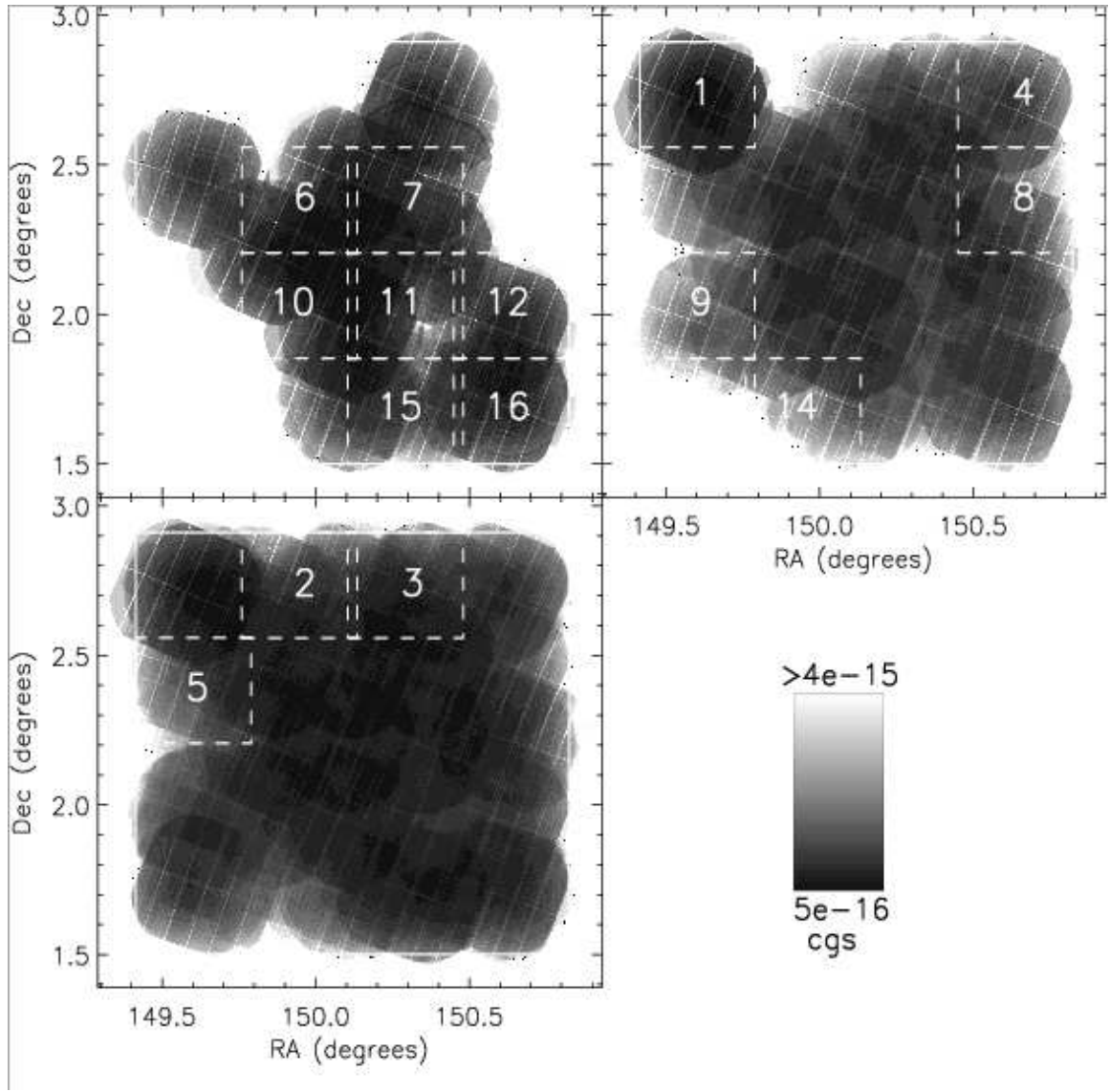}
\figcaption{Maps of X-ray sensitivity for each of the three years of
IMACS observing.  The top left panel shows the XMM depth and IMACS
pointings for the first year, the top right shows the second year
cumulative depth and pointings, and the lower left shows the third
year cumulative depth and pointings.  Since the XMM observations were
ongoing during the spectroscopy campaign, we chose each year's IMACS
pointings from the regions of greatest XMM uniformity and depth
lacking previous spectroscopic observations.
\label{fig:xmmsens}}
\end{figure}

\begin{figure}
\plotone{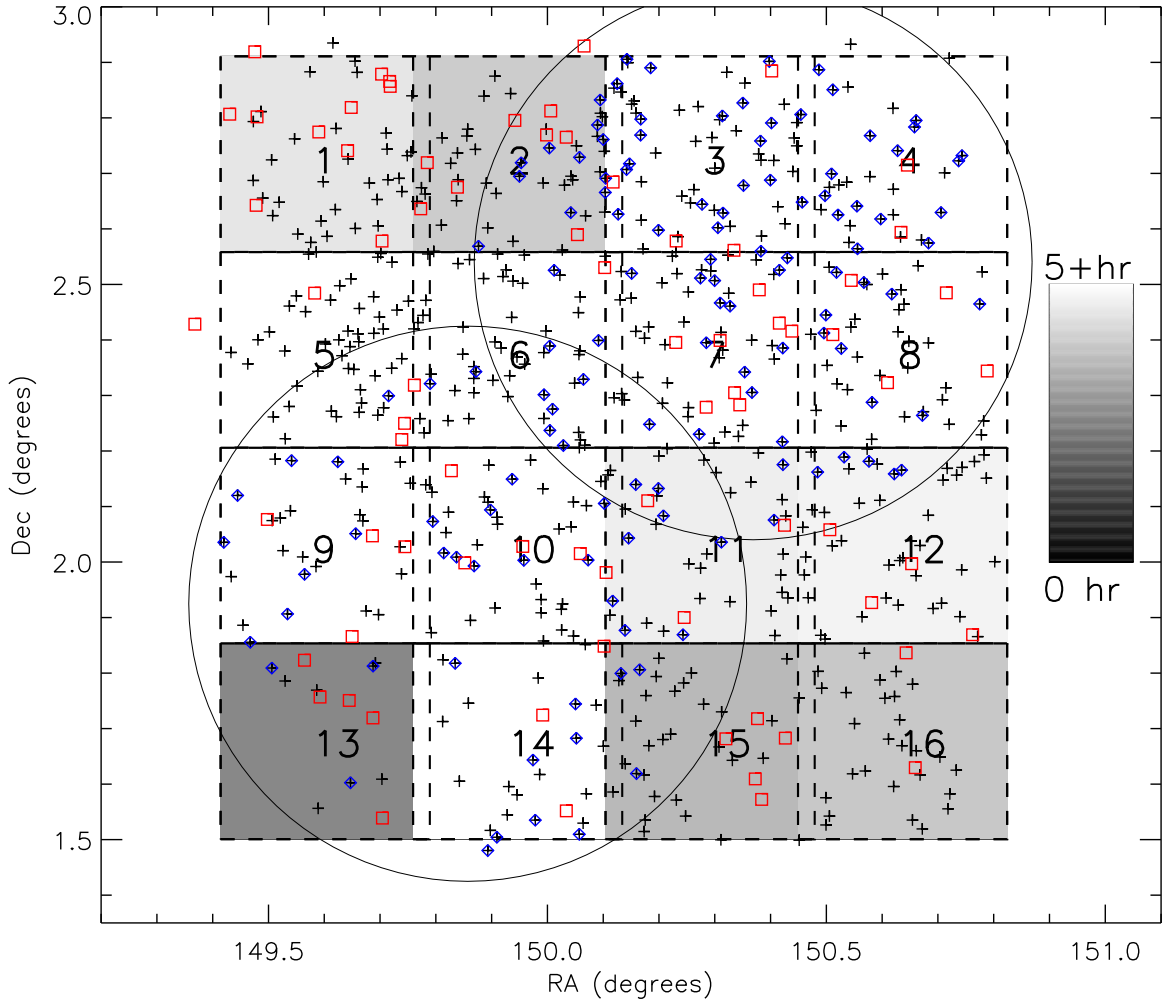}
\figcaption{Spectroscopic observations of the 2 deg$^2$ COSMOS area.
X-ray targets with IMACS spectra are shown as crosses, those with MMT
spectra are diamonds, and those with SDSS spectra are squares.  The 16
tiled IMACS pointings are shown as boxes of $22{\arcmin}30{\arcsec}
\times 21{\arcmin}10{\arcsec}$ and are shaded according to their
exposure time.  The two 1-deg diameter MMT pointings are shown as
circles.  COSMOS also includes deeper Chandra coverage, not used here,
over the field's central square degree (fields 6, 7, 10, and 11, with
portions of the other 8 fields).
\label{fig:field}}
\end{figure}

\begin{figure}
\plotone{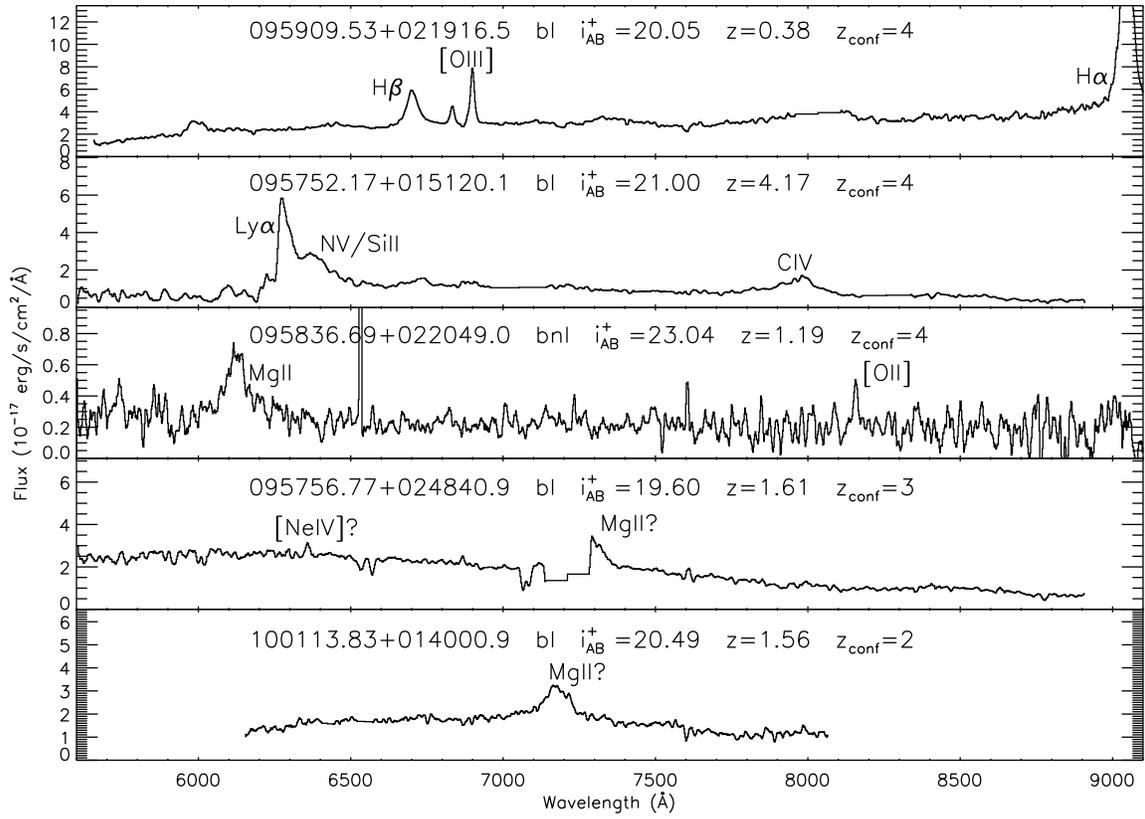}
\figcaption{Five examples of IMACS spectra with broad emission lines.
The dominant line species are labeled in each spectrum and bad pixels
are omitted.  The first two objects are Type 1 AGN (``bl'') with the
highest redshift confidences, the third is a high-confidence AGN with
both narrow and broad emission (``bnl''), and the bottom two are Type
1 AGN (``bl'') with uncertain redshifts.  We discuss these objects in
\S 2.2.
\label{fig:spec1}}
\end{figure}

\begin{figure}
\plotone{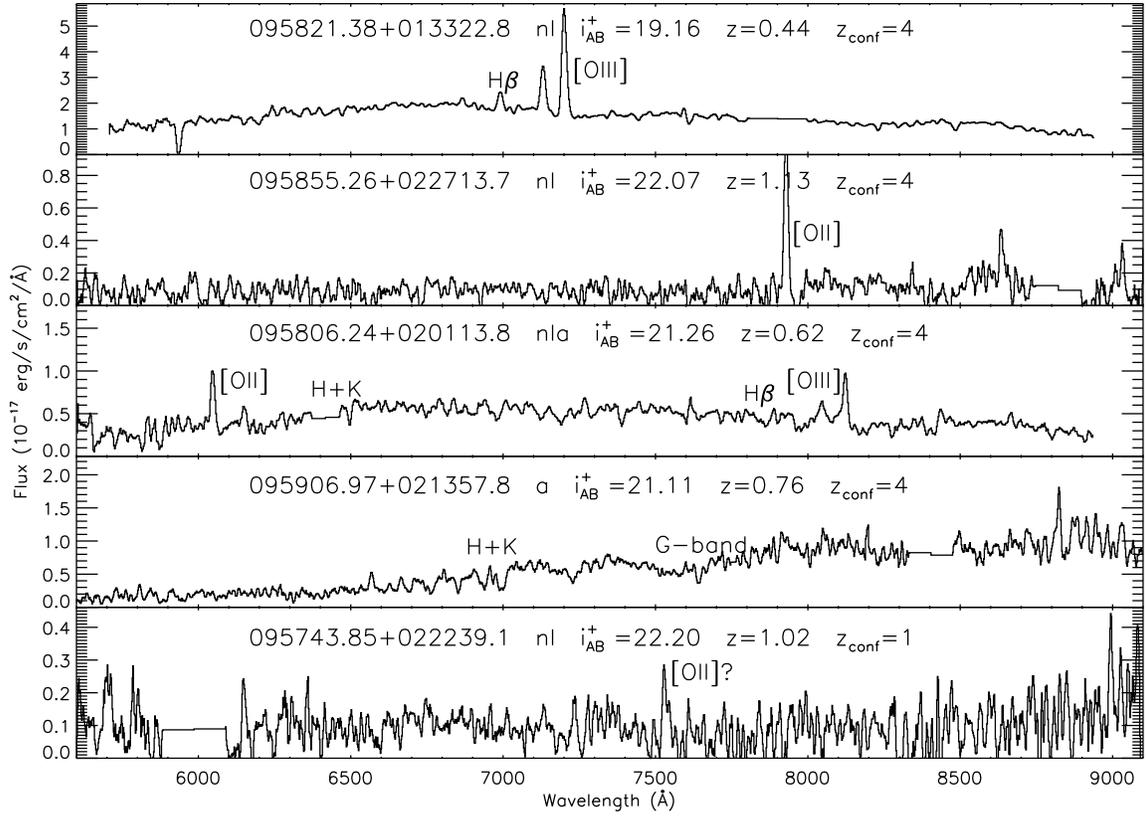}
\figcaption{Five more examples of IMACS spectra, including four
targets with narrow emission lines and one absorption line galaxy.
The prominent absorption and emission features are labeled.  The third
target is a hybrid ``nla'' object with both narrow emission and
absorption lines.  The first four objects have the highest redshift
confidence, while the bottom target has an extremely uncertain
redshift, calculated from a single emission line which may be solely
due to noise.  The first is a starburst galaxy by its X-ray emission,
while the second, third, and fourth spectra are all AGN which meet
both of the X-ray criteria of \S 3.2.  We discuss these objects in \S
2.2.
\label{fig:spec2}}
\end{figure}

\begin{figure}
\plotone{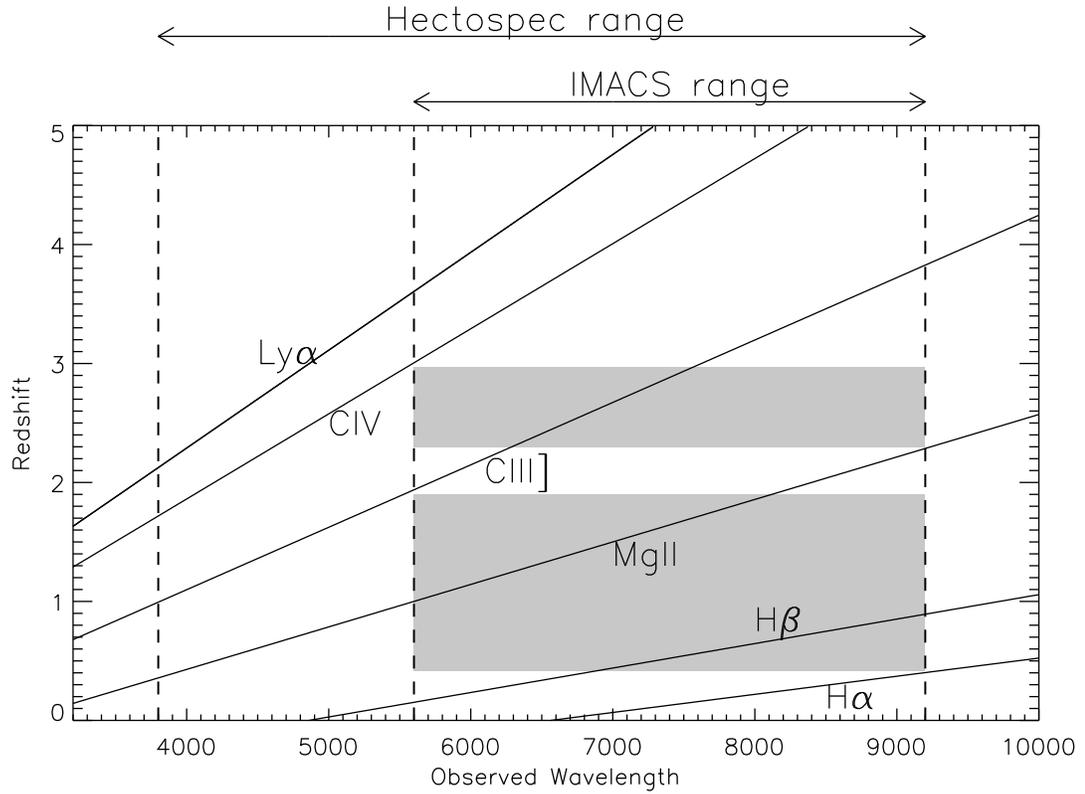}
\figcaption{The observed wavelengths of prominent broad emission lines
with redshift.  The spectral ranges of MMT/Hectospec and
Magellan/IMACS are shown at the top.  The broad emission lines
observed at a given redshift can be found by drawing a horizontal line
between the wavelength limits: the solid lines of broad emission peak
intersecting that redshift line would be present in the spectrum.  The
narrow wavelength coverage of IMACS means that only one broad line is
present in the shaded redshift ranges $0.4<z<1.9$ and $2.3<z<2.9$, so
that spectra with low S/N may be assigned $z_{\rm conf}=2$ because
they have degenerate redshift solutions.  The extended wavelength
coverage of Hectospec allows us to resolve the degeneracies and assign
$z_{\rm conf}=4.$
\label{fig:linefig}}
\end{figure}

\begin{figure}
\plotone{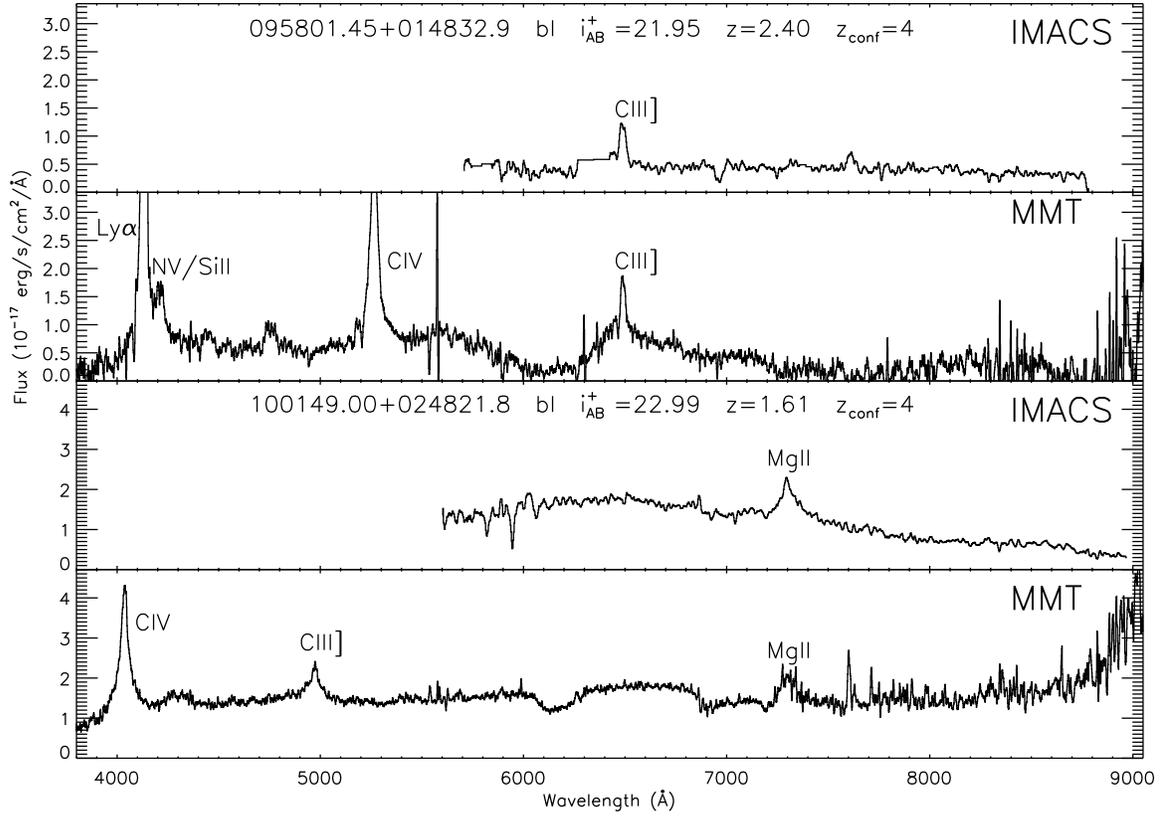} 
\figcaption{Two X-ray targets with both IMACS and Hectospec spectra.
In both cases, the IMACS wavelength range only includes one broad
emission line and so has a degenerate redshift solution.  The
additional blue Hectospec coverage resolves the degeneracy and allows
us to assign these objects $z_{\rm conf}=4$.  The fourth panel shows
that our Hectospec flux calibration can cause errors in spectral shape
at red wavelengths, although this does not affect our redshift
solutions.
\label{fig:imacsmmt}}
\end{figure}

\begin{figure}
\plotone{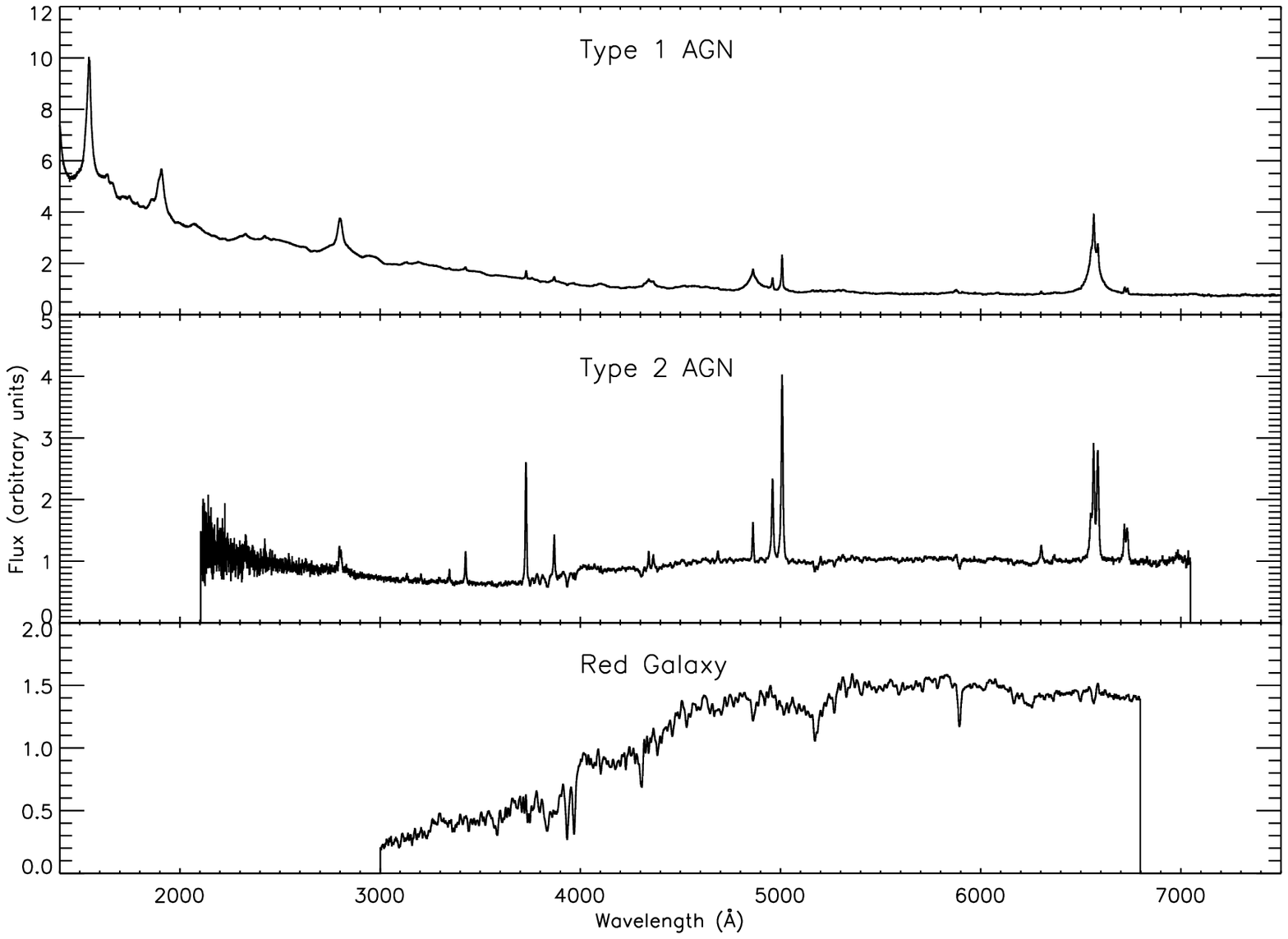}
\figcaption{The three templates used in the classification and
redshift determination scheme.  The broad line AGN template is the
SDSS quasar composite of \citet{q1template}, the narrow emission line
template is the SDSS Type II AGN composite of \citet{q2template}, and
the absorption line red galaxy template is the composite of the SDSS
red galaxy sample \citep{etemplate}.  The wavelength coverages of the
templates were sufficient for the entire redshift range (and the
corresponding observed wavelength ranges) of the sample.
\label{fig:templates}}
\end{figure}

\begin{figure}
\plotone{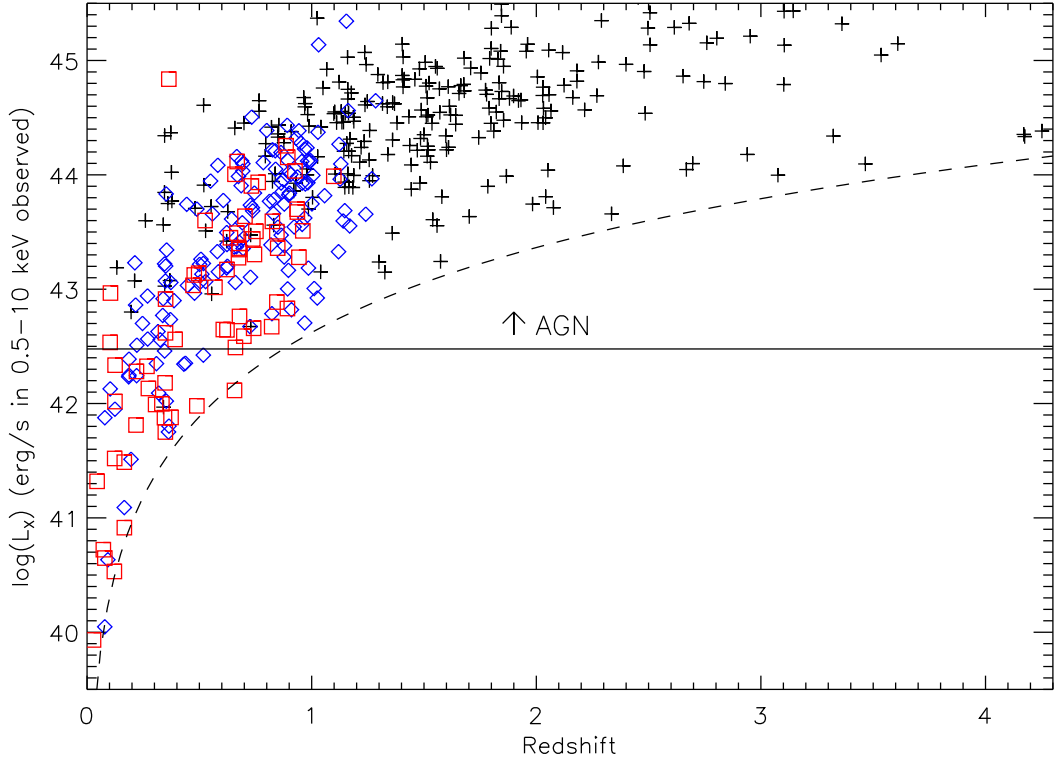}
\figcaption{The observed 0.5-10 keV X-ray luminosities for $z_{\rm
conf} \ge 3$ objects with redshift.  We label objects classified as
``bl'' (Type 1 AGN) with black crosses, ``nl'' with blue diamonds, and
``a'' with red squares.  The dashed line shows the survey's limiting
luminosity from the XMM flux limit.  The AGN luminosity cut of
$L_{0.5-10 \rm keV} = 3 \times 10^{42}$ is drawn as a solid line: all
``nl'' and ``a'' above this line are AGN.  Objects below this line,
however, are not necessarily inactive: two Type 1 AGN are less
luminous, and the luminosity limit is conservative.
\label{fig:xraylum}}
\end{figure}

\begin{figure}
\plotone{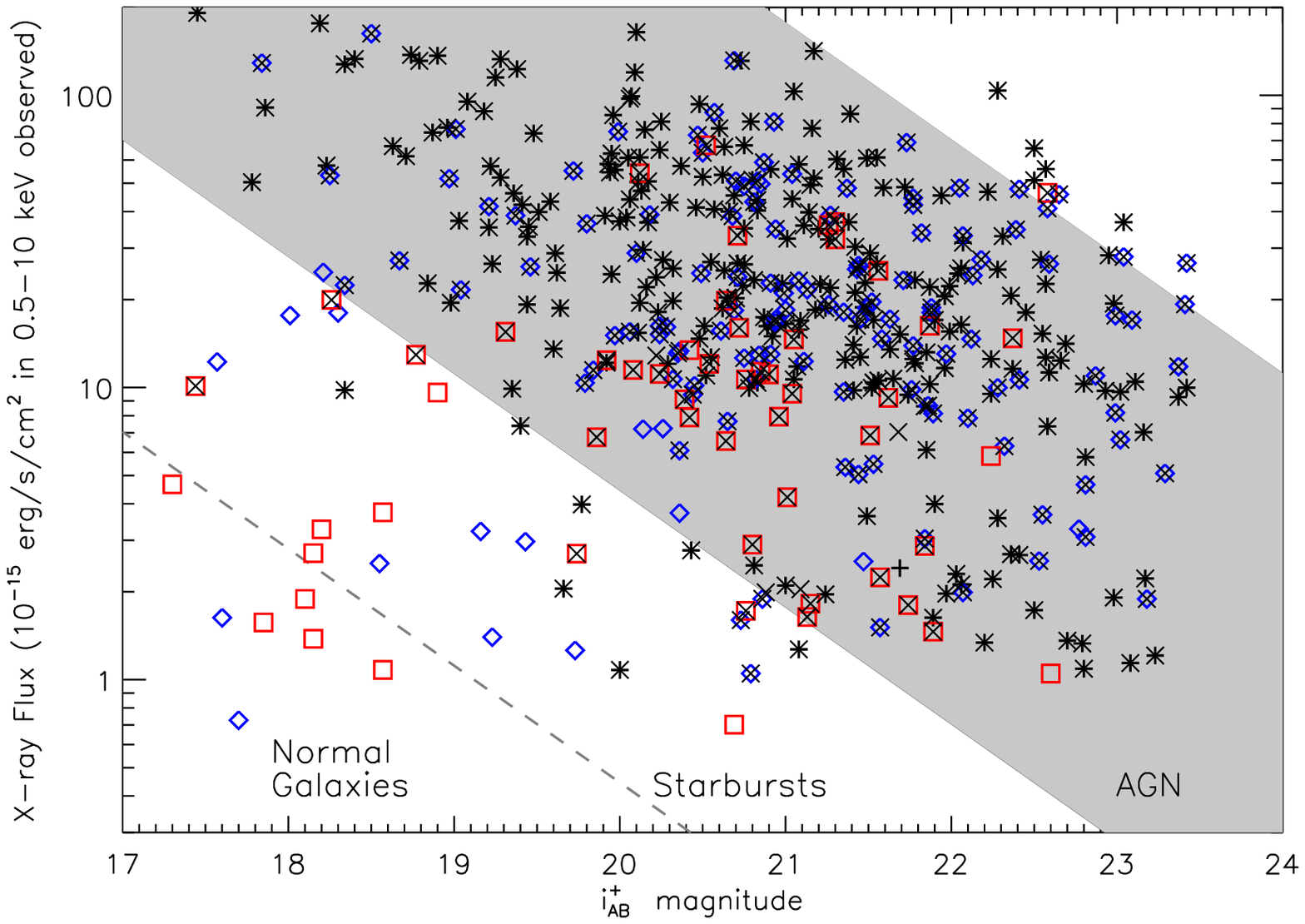}
\figcaption{The X-ray flux vs. the $i_{\rm AB}^+$ magnitude for the
$z_{\rm conf} \ge 3$ objects.  The AGN locus of $-1 \le \log{f_X/f_O}
\le 1$ \citep[][, see also Equation 2]{mac88} is shown by the gray
shaded region, along with approximate boundary between quiescent and
star-forming galaxies at $\log{f_X/f_O} = -2$ \citep{bau04}.  Black
crosses are targets classified ``bl'' (Type 1 AGN), blue diamonds are
``nl,'' and red squares are ``a'' objects.  We additionally mark all
targets of $L_{0.5-10 \rm keV} > 3 \times 10^{42}$ with black x's.  We
consider targets either in the AGN locus or with $L_X > 3 \times
10^{42}$ to be AGN: this includes all of the ``bl'' spectra and all
but 53 of the ``nl'' and ``a'' spectra.
\label{fig:xraylocus}}
\end{figure}

\begin{figure}
\plotone{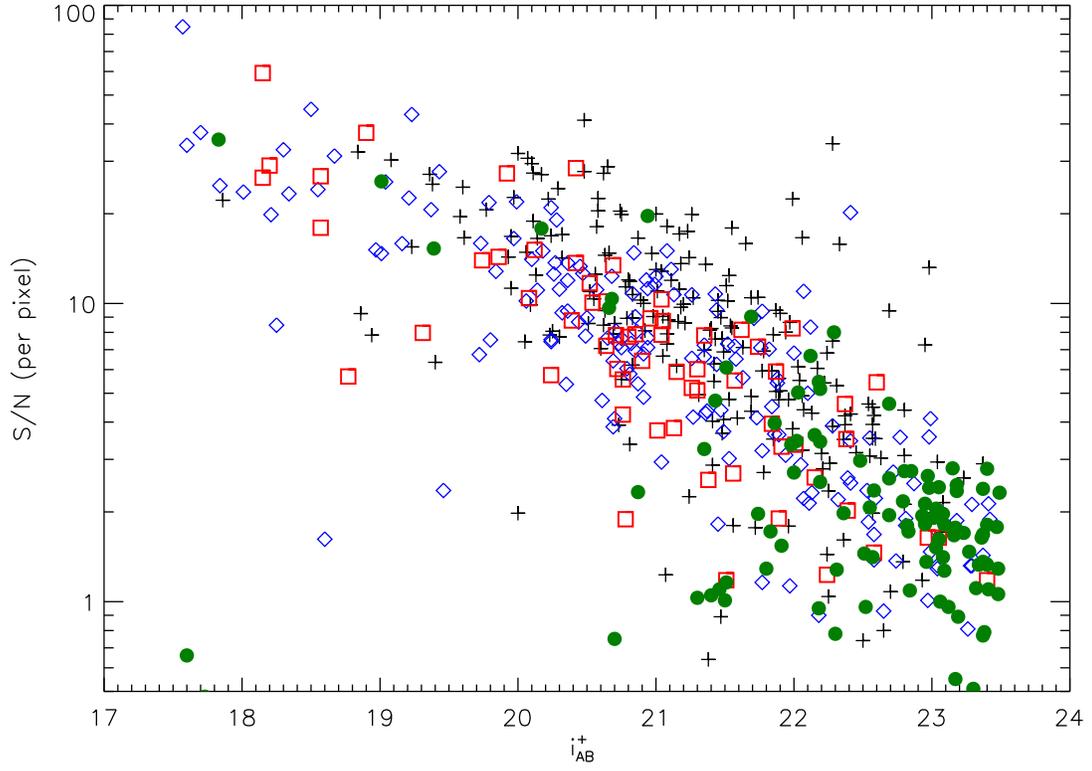}
\figcaption{The signal-to-noise (S/N) and optical $i_{\rm AB}^+$
magnitudes for the X-ray targets.  Crosses indicate broad emission
line spectra, diamonds are narrow emission line spectra, squares are
absorption line spectra, and filled circles are unclassified objects.
The S/N and optical magnitude are correlated, with scatter from
varying conditions over three years of observing.  The number of
unidentified objects increases greatly at $i_{\rm AB}^+>22$, although
we still identify emission line spectra at the faintest magnitudes.
\label{fig:sni}}
\end{figure}

\begin{figure}
\plotone{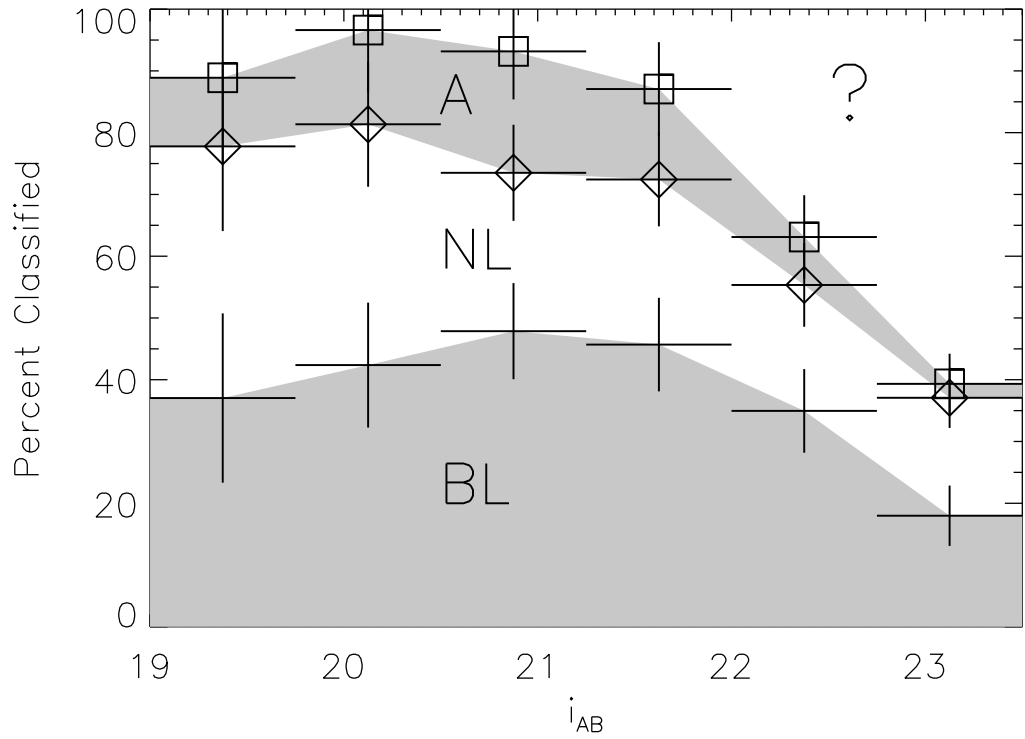}
\figcaption{Cumulative completeness by classification type versus
optical $i_{\rm AB}^+$ magnitude.  Each region (shaded or unshaded)
indicates the relative fraction of broad emission line objects
(``bl''), narrow emission line objects (``nl''), and absorption line
(``a'') spectra.  Targets with too low S/N to venture a classification
are represented in the upper ``?'' region.  Error bars on the points
above each region are calculated assuming that, in each magnitude bin,
both the number of each class and the total number have associated
Poisson counting errors.  The total completeness for each
classification is $\sim$90\% to $i_{\rm AB}^+ \le 22$, although we can
correct for the incompleteness of each spectral type to $i_{\rm AB}^+
\le 23$.
\label{fig:typesi}}
\end{figure}

\begin{figure}
\plotone{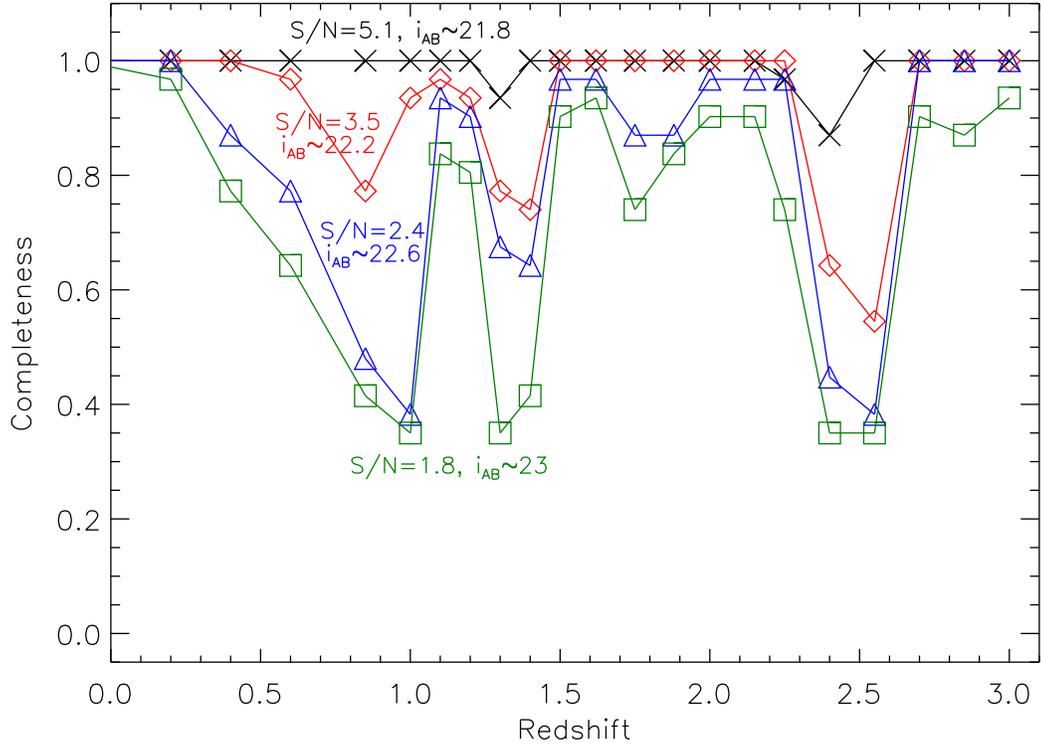}
\figcaption{Our estimated completeness to assigning $z_{\rm conf} \ge
3$ redshifts for broad emission line spectra (Type 1 AGN).  We used
Monte Carlo simulations with 20 different spectra with
Gaussian-distributed noise for each of 4 values of S/N and 20 redshift
bins.  Each point represents the number of simulated spectra assigned
$z_{\rm conf} \ge 3$, with an additional 36\% of the bad $z_{\rm
conf}<3$ simulated spectra based on the partial MMT/Hectospec coverage
(since 104/288 observed Type 1 AGN had supplemental Hectospec
spectra).  Each signal-to-noise is associated with an $i_{\rm AB}^+$
magnitude according to the median values in Figure \ref{fig:sni}.  The
redshift ranges of lowest completeness correspond to observed
wavelength ranges with only one emission line, as detailed in \S 4.3.
Almost all of the simulated spectra to which we are not complete are
identified as ``bl'' objects but with degenerate spectra.
\label{fig:q1complete}}
\end{figure}

\begin{figure}
\plotone{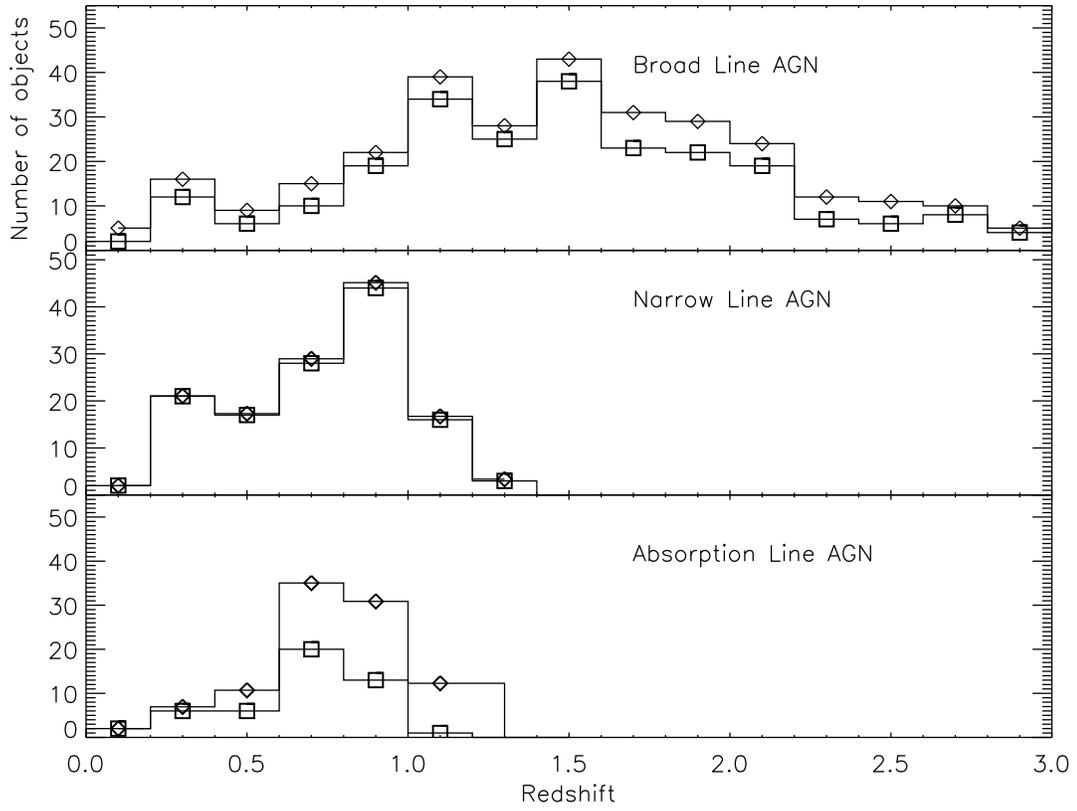}
\figcaption{The redshift distributions of broad line (``bl'') AGN,
narrow emission line (``nl'') spectra, and absorption line (``a'')
galaxies.  The raw distributions are shown with squares, while the
distributions adjusted for the incompleteness (see \S 4.3 \& 4.4 and
Figure \ref{fig:badphotz}) are shown with diamonds.  We do not 
correct the ``nl'' and ``a'' types where there is no spectroscopic 
data at $z>1.4$.
\label{fig:zhist}}
\end{figure}

\begin{figure}
\plotone{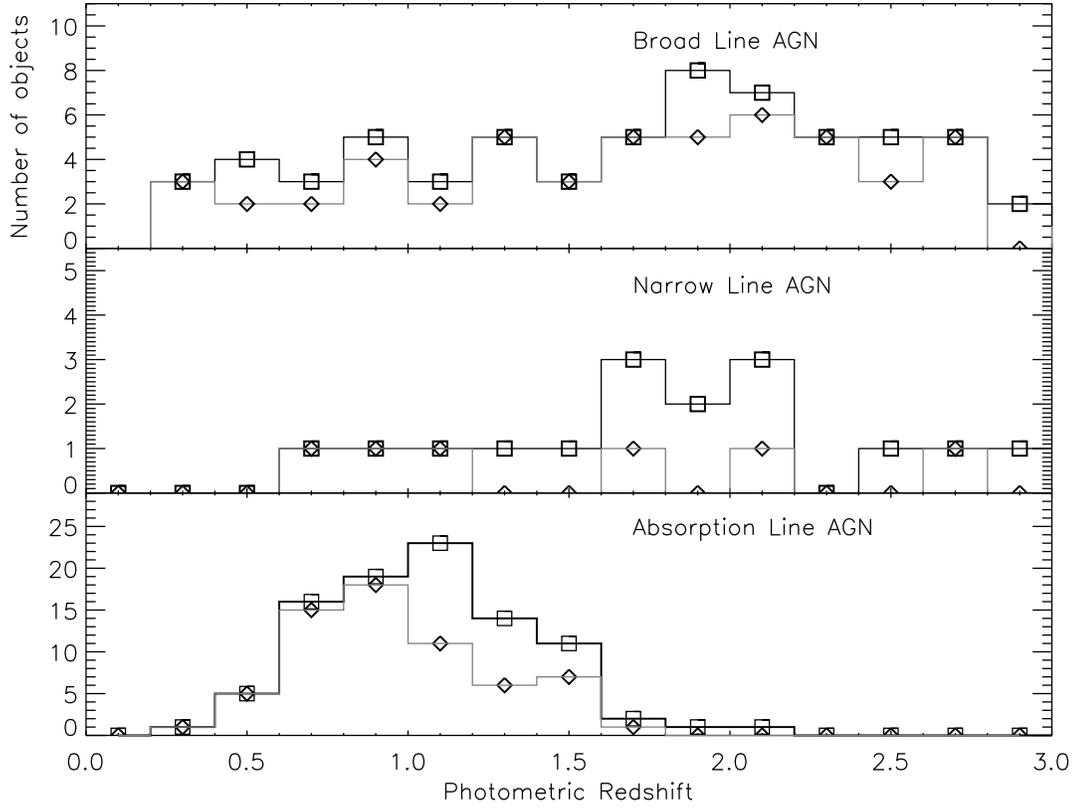}
\figcaption{The photometric redshifts for the spectra without
high-confidence redshifts.  Square symbols show all 192 objects, and
diamonds show the 146 $i_{\rm AB}^+<23$ objects.  The spectral type
for these objects comes from the template used for the photometric
redshift, with ``a'' and ``nl'' fractions estimated as described in \S
4.4.  We use the photometric redshifts and classifications to
characterize the complete $i_{\rm AB}^+<23$ X-ray AGN sample.
\label{fig:badphotz}}
\end{figure}

\begin{figure}
\plotone{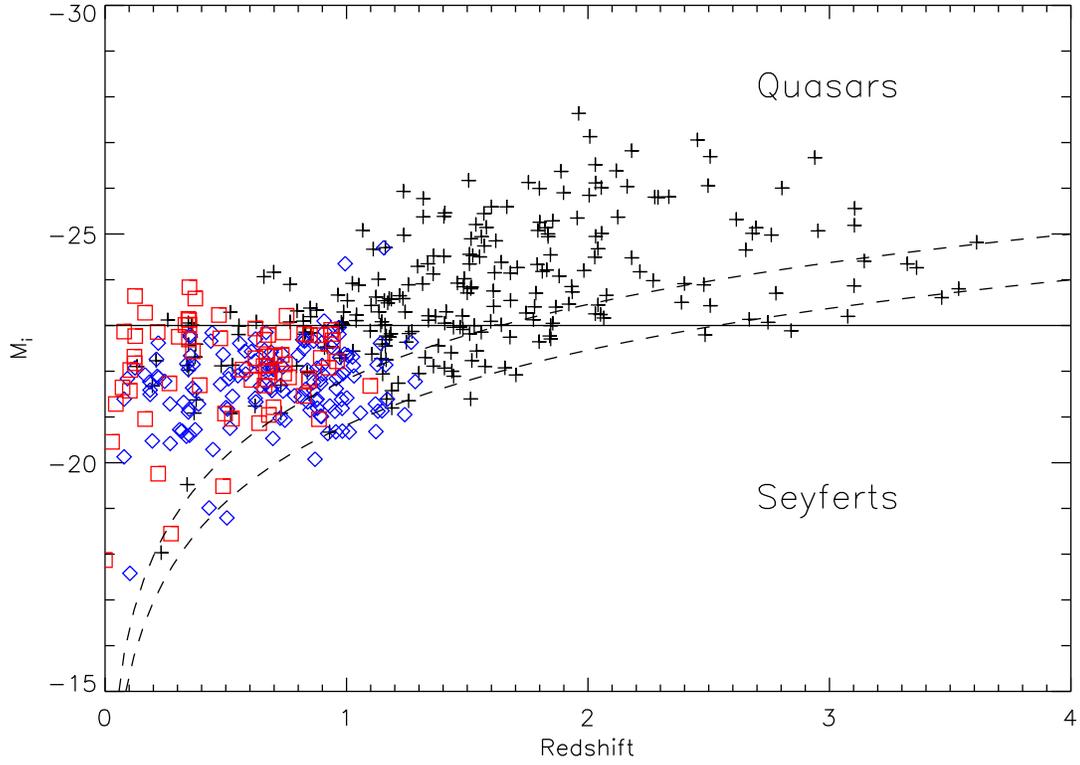}
\figcaption{The absolute magnitude of the AGN (meeting the X-ray
criteria of $-1 \le \log{f_X/f_O} \le 1$ or $L_{0.5-10 \rm keV} > 3
\times 10^{42}$ ) with redshift.  The solid line indicates an
arbitrary quasar/Seyfert boundary at $M_i=-23$, and the dashed lines
indicate fluxes of $i_{\rm AB}^+=22$ and $i_{\rm AB}^+=23$.  Symbols
for object types are as in Figures \ref{fig:sni} and \ref{fig:typesi},
with crosses for ``bl'' or ``bnl'' objects (Type 1 AGN), diamonds for
``nl'' and ``nla'' objects (Type 2 AGN), and squares for ``a'' objects
(optically obscured AGN).
\label{fig:absmag}}
\end{figure}

\begin{figure}
\plotone{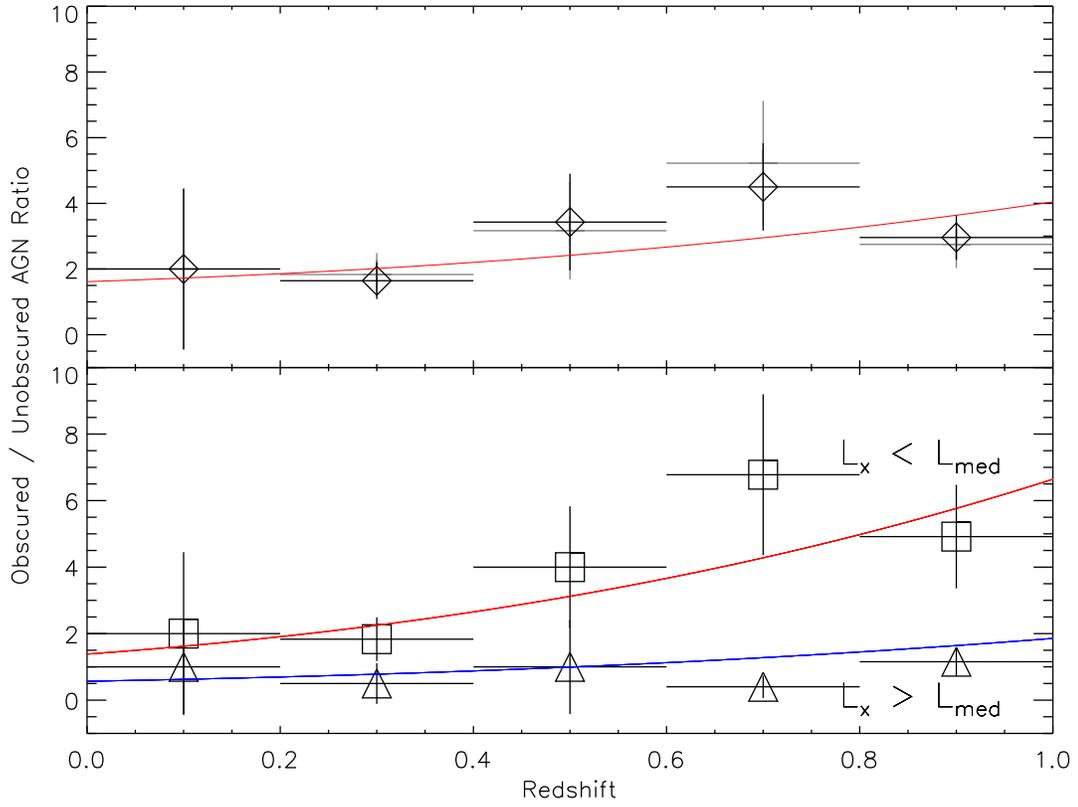}
\figcaption{The ratio of obscured to unobscured AGN with redshift.  We
define obscured AGN as spectra with narrow emission or absorption
lines (``nl,'' ``a,'', and ``nla'') that meet the X-ray AGN criteria
of \S 3.2, while unobscured AGN are all broad-line (``bl'' and
``bnl'') spectra.  In the top panel the raw fractions are shown in
gray, while the corrected fractions based on the incompleteness
(characterized in \S 4.3 \& 4.4) are shown by the black diamonds.  The
bottom panel shows the ratios for AGN fainter and brighter than the
median luminosity $L_{\rm med}=1.32 \times 10^{44}$ cgs, with the
$L_{0.5-10~ \rm keV}<L_{\rm med}$ ratio as squares, and the
$L_{0.5-10~ \rm keV}>L_{\rm med}$ ratio as triangles.  The errors
associated with each point assume that the numbers of objects observed
in each redshift bin are Poissonian.  Logistic regression analysis
shows that the ratio of obscured to unobscured increases with redshift
and decreases with luminosity, as shown by the best-fit power-laws.
\label{fig:obscurfrac}}
\end{figure}

\end{center}

\end{document}